\documentclass[5p]{elsarticle}

\usepackage{amssymb}
\usepackage{amsmath}
\usepackage{xspace}
\usepackage{textcomp} 
\usepackage{subfig}

\newcommand{\bigON}{$\mathcal{O}(N)$}
\newcommand{\bigOsq}{$\mathcal{O}(N^2)$}
\newcommand{\bibee}{\textsc{bibee}\xspace}
\newcommand{\cfa}{\textsc{cfa}\xspace}
\newcommand{\cdk}{\textsc{cdk}\xspace}
\newcommand{\nvidia}{n\textsc{vidia}\xspace}

\newcommand{\gmres}{\textsc{gmres}\xspace}
\newcommand{\fmm}{\textsc{fmm}\xspace}
\newcommand{\gpu}{\textsc{gpu}}
\newcommand{\cpu}{\textsc{cpu}}
\newcommand{\cuda}{\textsc{cuda}\xspace}
\newcommand{\bem}{\textsc{bem}\xspace}
\newcommand{\uKern}{\mathbb{K}}
\newcommand{\ME}{\textsc{me}}
\newcommand{\LE}{\textsc{le}}
\newcommand{\PM}{\textsc{p}2\textsc{m}\xspace} 
\newcommand{\MM}{\textsc{m}2\textsc{m}\xspace} 
\newcommand{\ML}{\textsc{m}2\textsc{l}\xspace} 
\newcommand{\LL}{\textsc{l}2\textsc{l}\xspace}  
\newcommand{\LP}{\textsc{l}2\textsc{p}\xspace}  
\newcommand{\PP}{\textsc{p}2\textsc{p}\xspace}  
\newcommand{\petfmm}{\textsc{p}et\textsc{fmm}\xspace}

\usepackage[
    pdftex,
    pdftitle={Biomolecular electrostatics using a fast multipole BEM on up to 512 GPUs and a billion unknowns},
    pdfauthor={Rio Yokota, Jaydeep P. Bardhan, Matthew G.~Knepley, L.~A. Barba},
    pdfpagemode={UseOutlines},
    bookmarks, bookmarksopen,bookmarksnumbered={True},
    colorlinks, linkcolor={blue},citecolor={blue},urlcolor={blue}
    ]{hyperref}
\makeatletter
\def\url@leostyle{%
  \@ifundefined{selectfont}{\def\UrlFont{\sf}}{\def\UrlFont{\small\ttfamily}}}
\makeatother
\begin{document}

\begin{frontmatter}

\title{Biomolecular electrostatics using a fast multipole BEM on up to 512 GPU and a billion unknowns}


\author[lab]{Rio Yokota}
\ead{yokota@bu.edu}

\author[jpb]{Jaydeep P. Bardhan}
\ead{jbardhan@alum.mit.edu}

\author[mgk]{Matthew G. Knepley}
\ead{knepley@ci.uchicago.edu}

\author[lab]{L.~A.~Barba\corref{lab2}}
\ead{labarba@bu.edu}

\author[th]{Tsuyoshi Hamada}
\ead{hamada@nacc.nagasaki-u.ac.jp}

\address[lab]{Department of Mechanical Engineering, Boston University, Boston MA 02215}
\address[jpb]{Dept.\ of Molecular Biophysics and Physiology, Rush University Medical Center, Chicago, IL 60612}
\cortext[lab2]{Correspondence to:  110 Cummington St, Boston MA 02215, (617) 353-3883, \href{mailto:labarba@bu.edu}{labarba@bu.edu}}
\address[mgk]{Computation Institute, University of Chicago, Chicago, IL 60637}
\address[th]{Nagasaki University, Advanced Computing Center (NACC), Nagasaki, Japan}

\begin{abstract}
We present teraflop-scale calculations of biomolecular electrostatics
enabled by the combination of algorithmic and hardware acceleration.
The algorithmic acceleration is achieved with the fast multipole
method (\fmm) in conjunction with a boundary element method (\bem)
formulation of the continuum electrostatic model, as well as the
\bibee approximation to \bem.  The hardware acceleration is achieved
through graphics processors, \gpu s.  We demonstrate the power of our
algorithms and software for the calculation of the electrostatic
interactions between biological molecules in solution.  The
applications demonstrated include the electrostatics of protein--drug
binding and several multi-million atom systems consisting of hundreds
to thousands of copies of lysozyme molecules.  The parallel
scalability of the software was studied in a cluster at the Nagasaki
Advanced Computing Center, using 128 nodes, each with 4 \gpu s.
Delicate tuning has resulted in strong scaling with parallel
efficiency of 0.8 for 256 and $0.5$ for $512$ GPUs. The largest application run, with
over 20 million atoms and one billion unknowns, required only one
minute on 512 \gpu s. We are currently adapting our \bem software to
solve the linearized Poisson--Boltzmann equation for dilute ionic
solutions, and it is also designed to be flexible enough to be
extended for a variety of integral equation problems, ranging from
Poisson problems to Helmholtz problems in electromagnetics and
acoustics to high Reynolds number flow.
\end{abstract}

\end{frontmatter}


\section{Introduction}

Electrostatic interactions play an essential role in the structure and
function of biomolecules (proteins, \textsc{dna}, cell membranes,
\emph{etc.})  \cite{SharpHonig1990,WarshelETal2006}.  One of the most
challenging aspects for understanding these interactions is the fact
that biologically active molecules are almost always in
solution---that is, they are surrounded by water molecules and
dissolved ions.  These solvent molecules add many thousands or even
millions more degrees of freedom to any theoretical study, many of
which are of only secondary importance for investigations of interest.
Classical molecular dynamics (\textsc{md}) methods, implemented in
software libraries such as \textsc{charmm}~\cite{BrooksETal1983} and
\textsc{namd}~\cite{PhillipsETal2005}, use all-atom representations of
the biomolecule and solvent, and compute the trajectories of every
atom over time by following Newton's equations of motion.  \textsc{md}
methods are the most detailed approach to studying biomolecular
systems, but computing an ``average'' electrostatic energy in such
systems can be extremely expensive, even when one uses efficient
methods for the long-range electrostatics between all the atoms, such
as particle-mesh Ewald~\cite{DardenETal1993} or multilevel summation
\cite{SkeelETal2002}.  For many studies, a faster method based on
adequate approximations is a necessity.

In contrast with all-atom \textsc{md} computations, one can model the
electrostatic interactions in solvated molecules using a continuum
representation.  Such a model is based on assuming that the molecules and the
solvent can be treated as continuous dielectric media with different
dielectric constants, low and high, respectively.  Inside the
biomolecule, in addition, point charges are arranged explicitly at the
atomic positions. Thus, the electrostatic potential can be described
by a Poisson equation, which for general shapes of the dielectric
boundary has to be solved numerically.  Accounting for ionic charge
distributions in the solvent introduces some extra complications, as
the ion locations depend on the combined effect of all charges,
dielectric distributions, and the ions themselves.  Making the
assumption that the average electrostatic potential multiplied by the
charge of the ion determines the mean force acting on the ion
particle, a Poisson--Boltzmann model for biomolecular systems is
obtained~\cite{SharpHonig1990}.  Standard numerical approaches such as
finite-difference methods and finite-element methods can be used to
solve the Poisson or Poisson--Boltzmann
equations~\cite{WarwickerWatson1982,GilsonETal1985,BakerETal2001}, but
there are several challenges that must be overcome.  These include the
difficulty of mapping an irregular molecular surface to a volumetric
mesh, representing the source distribution as a set of discrete point
charges, and convergence issues associated with dielectric
discontinuities.  The development of various strategies to mitigate
these problems~(see, \emph{e.g.},
\cite{BruccoleriETal1997,GengYuWei2007}) and the availability of
scalable, open-source software such as
\textsc{apbs}~\cite{BakerETal2001} have helped make the continuum
model a popular approach for studying molecular electrostatics.

An alternative approach to solving the Poisson equation directly is to
use a boundary-integral formulation and the boundary-element method,
\bem, to determine the induced charge distribution on the molecular
surface which accounts for the change in polarization charge across
the dielectric boundary.  One significant advantage of a
\bem\ formulation is the fact that the discretization is performed
over the surface of the biomolecule, rather than the three-dimensional
volumetric region occupied by the molecule and solvent.  Accurate
solution of the linearized form of the Poisson--Boltzmann equation is
also possible with \bem\ using various specialized
techniques~\cite{JufferETal1991, Zhou1993,LuETal2006,
  AltmanBardhanWhiteTidor09}.  The main challenge in the \bem\ is the
computational cost of finding the induced-charge distribution, which
is obtained by solving a linear system in which the matrix is dense
(in contrast to the sparse linear systems associated with
finite-difference and finite-element methods).  To greatly reduce the
expense of solving such a linear system using Krylov iterative methods
such as \gmres~\cite{SaadSchultz1986}, the fast multipole method
(\fmm)~\cite{Rokhlin1983,GreengardRokhlin1987, NaborsWhite1991} can be
used to calculate the dense matrix-vector product needed in the
iterative solver in $\mathcal{O}(N)$ operations
\cite{BharadwajETal1995}.  In this way, the \fmm\ algorithm and
similar approaches~\cite{PhillipsWhite1997,AltmanETal2006} can enable
calculations with hundreds of thousands or even millions of degrees of
freedom~(\emph{e.g.} \cite{BharadwajETal1995,KuoETal2002,LuETal2006}).

In this paper, we demonstrate a fast molecular electrostatics
application using a \bem\ formulation of the continuum model. The
application is accelerated both at the algorithmic level and by
hardware, achieving an unprecedented capacity for simulating large
biomolecular systems.  Our largest example models a collection of
biomolecules, totalling more than 20 million atoms, for which the \bem
problem has more than \emph{one billion unknowns}. It was accomplished using a cluster of 128 nodes with a total of 512 \gpu s and required approximately one minute to
complete. Thus, the electrostatic interactions between large proteins
and molecular machines with tens to hundreds of thousands of atoms can
be simulated in a few minutes on a single \gpu\ or small
\gpu\ cluster.  This computational performance transforms the
landscape for theoretical  investigations of biomolecular
electrostatics such that the new limiting factor is the generation of
suitable meshes for the \bem, rather than fast solver approaches.  We
wish to emphasize that we are referring to calculations of
static structures (for example, as in the post-processing stages of
\textsc{md}-based estimates of binding free energies using \textsc{mm-pbsa}
methods~\cite{MassovaKollman2000}) rather than actual dynamics.  Substantial
hurdles remain before \bem electrostatics can be practical for
implicit-solvent dynamics, including both meshing and the calculation
of suitable forces~\cite{LuMcCammon2007,StorkTavan2007}.  Our solver is currently
experimental, but already available publicly via the repository of the
open-source \petfmm library\footnote{To obtain the software and
  documentation, follow the links provided in
  \href{http://barbagroup.bu.edu/}{http://barbagroup.bu.edu/}}, which
provides a parallel implementation of the \fmm with dynamic load
balancing \cite{CruzKnepleyBarba2010}.  The \gpu\ kernel
implementations for the \gpu\ architecture are derived from the 2009
Gordon Bell prize-winning \fmm\ code presented in
\cite{HamadaNarumiYokotaYasuokaNitadoriTaiji09}.

There have been some notable recent publications reporting
\gpu-accelerated algorithms related to our present contribution.  The
multilevel summation method \cite{SkeelETal2002} for calculating
electrostatic potentials is implemented for the \gpu\ architecture in
\cite{HardyStoneSchulten2009}.  In this algorithm, the short-range
interactions are equivalent to the particle-to-particle (\PP)
operations in the \fmm, while the long-range forces are obtained from
hierarchical interpolations; both the short-range interactions, and
the dominant part of the long-range ones were implemented as
\gpu\ kernels, achieving an overall speed-up of $26\times$ over the
\cpu\ version.  However, we note that the algorithm is completely
dominated by the short-range pair-wise interactions, which on the
\cpu\ take 90\% of the runtime and on the \gpu\ take 75\%, as shown in
Table 1 of \cite{HardyStoneSchulten2009}.  In this respect, the
approach is quite different from the one followed with the \fmm, where
one aims for an optimal balance between the \bigOsq\ \PP and the
multipole-to-local (\ML) transformation (see Figure
\ref{fig:UpDown-sweeps} for definitions of the \fmm operations).  The
demonstration runs in \cite{HardyStoneSchulten2009} use a 1.5 million
atom water box, with a total runtime of 20 s on one \textsc{gtx} 280
\gpu.

Another related work to the present contribution was presented in
\cite{LashukETal2009}, where a variant of the \fmm called the
kernel-independent fast multipole method \cite{YingBirosZorin2004} is
implemented for multi-\gpu\ systems.  That algorithm and
implementation was demonstrated on an impressive 256 million
point-system, taking 2 s to compute the interaction on 256 \gpu s.
The test consisted of a uniform random distribution of particles in a
unit cube (tests with non-uniform distributions are also reported, but
on \cpu-only systems), and a consistent speed-up measure of $25\times$
is obtained from the \gpu\ on weak scaling tests.  This \gpu-accelerated kernel-independent \fmm has been recently used in an award-winning application to simulation of red-blood cells \cite{RahimianETal2010}.  Finally, in \cite{TakahashiHamada2009}
a \gpu-accelerated \bem\ for the Helmholtz equation is presented and
demonstrated up to one million unknowns.  However, this work does not
use algorithmic acceleration by means of the \fmm, performing instead
the \bigOsq\ method.

Even with linear-scaling algorithms such as multigrid
approaches for finite-element methods \cite{BakerETal2001} and \fmm
for \bem, the computational costs of solving the continuum
electrostatic model can be prohibitive, especially when the scientific
questions of interest require the simulation of many millions of
unknowns.  Such investigations include \textit{in silico} screening of
candidate drugs~\cite{LiuGrinterZou2009}, protein
design~\cite{GreenTidor2005,VizcarraMayo2005}, and implicit-solvent \textsc{md}, in
which one computes the trajectories of all protein atoms but replaces
the effect of solvent molecules with an appropriate potential of mean
force~\cite{GilsonMcCammonMadura1995}.  These applications have driven the development
of much faster approximate models such as the popular generalized-Born
(\textsc{gb}) models~\cite{StillETal1990} which are frequently used in
implicit-solvent molecular dynamics, \emph{e.g.}~\cite{FanETal2005}.
Recently, more innovative approaches have also been described, many of
which also employ \gpu s ~\cite{FenleyETal2008,AnandakrishnanETal2010}.  In the
present paper, we approximate the \bem solution using the new \bibee
(boundary-integral-based electrostatics estimation)
model~\cite{Bardhan08_BIBEE,BardhanKnepleyAnitescu09}, which can be
more than an order of magnitude faster than \bem simulation; thus,
\bibee is well-suited for rapid screening of candidate drug molecules
(a common application of solvers such as
\textsc{apbs}~\cite{BakerETal2001}), but like \bem it is not presently
applicable to computing dynamics.  Modern \textsc{gb} models and
implementations have been extensively parameterized and optimized,
giving them certain performance and accuracy advantages over early,
unoptimized implementations of \bibee when computing electrostatic
solvation free energies.  However, \bibee does offer several of its own
advantages, including the fact that it allows scientists to fully
leverage well-known numerical techniques such as \fmm, rather than
necessitating new implementation.  The ability to employ sophisticated
computational primitives such as \fmm as an algorithmic substrate is
particularly important in the face of rapidly evolving hardware
architectures.  Thus, there remains a healthy (and in our view,
productive) tension between efforts to develop general fast methods,
one of which we apply in this work, and efforts to specialize existing
fast methods for particular problems such as biomolecular
electrostatics~\cite{FenleyETal2008,AnandakrishnanETal2010}.

This contribution aims primarily at demonstrating the power of
multiplying speedups: fast linear-scaling algorithms, rigorous
approximation of the continuum dielectric model using \bibee, and
hardware acceleration with \gpu s.  The first wave of successful
applications of the \gpu\ in scientific computing was crowded with
highly parallel algorithms (like \textsc{md}) which fit well to the
hardware architecture.  As could be expected, it is generally much
more difficult to obtain high performance with the more elaborate
hierarchical and \bigON\ algorithms.  But it is in this combination
where leaps in computing capability are possible which are orders of
magnitude larger than what Moore's law would achieve in a given
period.  In the bioelectrostatics application, the multiplying
speedups of algorithm and hardware are further enhanced by a
mathematical model that does not lavish computational effort on
nonessential degrees of freedom (the solvent molecules).  Many
applications may still require detailed modeling at the microscale,
but where the continuum approach gives sufficient accuracy, the
methods and software of our contribution can enable high-impact
advances.

\section{Background on the models and algorithm}\label{s:background}

\subsection{Bioelectrostatics using the continuum model}

\begin{figure}
	\centering
	{\includegraphics[width=0.2\textwidth]{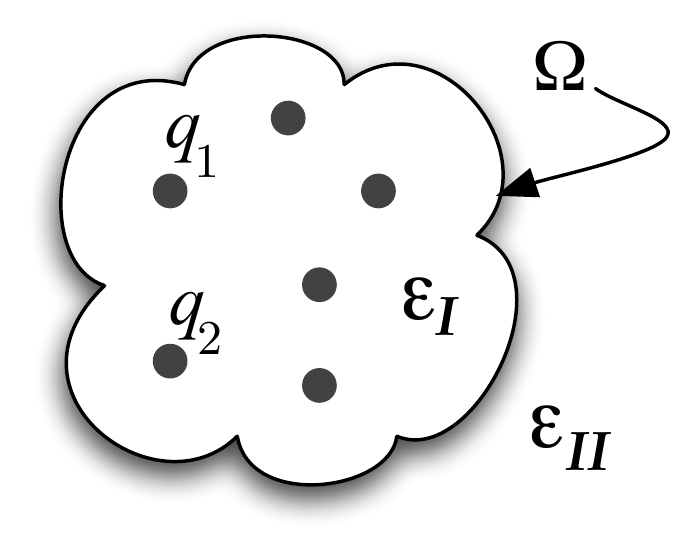}}
	\caption{In the continuum, implicit-solvent model for biomolecular electrostatics, the interior of the biomolecule is a dialectric with permittivity $\epsilon_I$ and free charges $q_{i}$, while the exterior, solvent-filled region is a dialectric material with permittivity $\epsilon_{II}$.}
	\label{fig:electrostatic-model}
\end{figure}

The continuum electrostatic model we treat in this work is a
mixed-dielectric Poisson problem, described as follows (see Figure
\ref{fig:electrostatic-model}).  The molecular interior, denoted as
region $I$, is treated as a homogeneous dielectric with permittivity
$\epsilon_I$; typical values are between 2 and 10~\cite{SharpHonig1990,SchutzWarshel2001}).
The molecular charge distribution is modeled as a set of $n_c$
discrete point charges located at the atom centers.  Denoting the
$i^{\text{th}}$ charge as having value $q_i$ and position $r_i$, the
electrostatic potential in this region satisfies a Poisson equation,
\begin{equation}
  \nabla^2 \varphi_I(r) = -\sum_{i=1}^{n_c} \frac{q_i}{  \epsilon_I} \delta(r - r_i).  
  \end{equation}

The water surrounding the molecule, corresponding to region $II$, is also modeled as a
homogeneous dielectric, of permittivity $\epsilon_{II}$ equal to that
of bulk water (approximately 80).  In this region, the Laplace equation
$\nabla^2 \varphi_{II}(r) = 0$ holds, because we assume that there are no
fixed or mobile charges.  At the dielectric boundary, $\Omega$, the
potentials and (in the absence of a free surface charge) the normal components of the electric displacement fields are
continuous:
\begin{eqnarray}
  \varphi_I(r_{\Omega}) &=& \varphi_{II}(r_{\Omega})\\
  \epsilon_I \frac{\partial \varphi_I}{\partial n}(r_{\Omega}) &=& \epsilon_{II} \frac{\partial \varphi_{II}}{\partial n}(r_{\Omega}).
\end{eqnarray}

\noindent Here, $n$ denotes the outward unit normal vector at
$r \in \Omega$, pointing into region $II$ from region $I$.  We define
the boundary by rolling a probe sphere (designed to mimic a water
molecule) over the union of spheres that represent the solute atoms,
a definition known as the solvent-excluded surface (also known as the
molecular surface)~\cite{Richards1977a,Connolly1983a}.

The solute charges polarize the solvent, which in turn creates a
\textit{reaction potential} in the solute.  In the mixed-dielectric
continuum model, the solvent polarization appears as a layer of
induced charge $\sigma(r)$ at the dielectric interface, where
$\sigma(r)$ satisfies the second-kind Fredholm boundary integral
equation~\cite{Rush66,Miertus1981,Shaw85,Bardhan2009}:

\begin{equation}
\noindent \left(1 -\frac{\epsilon_I}{\epsilon_{II}}\right)
\left(\frac{1}{4 \pi}\frac{\partial}{\partial n}\sum_{i=1}^{n_c}\frac{q_i}{||r-r_i||}
+  \frac{1}{4 \pi}\frac{\partial}{\partial n}\int_\Omega \frac{\sigma(r')}{ ||r-r'||}dA'
\right) 
= \sigma(r).
\label{eq:ASC}
\end{equation}

\medskip
The reaction potential in the solute is then just the Coulomb potential
induced by $\sigma(r)$, \emph{i.e.},
\begin{equation}
  \varphi^{\mathrm{REAC}}(r) = \int_\Omega \frac{\sigma(r')}{4 \pi || r - r' ||} dA',\label{eq:reac}
  \end{equation}
such that the total potential $\varphi_I(r)$ is the sum of
$\varphi^{\mathrm{REAC}}$ and the bare Coulomb potential induced by
the point charges.  The electrostatic energy associated with the
reaction potential represents the change in electrostatic energy
associated with transferring the given charge distribution and
molecular shape from a uniform low dielectric into the high dielectric
medium.  For this reason, the energy is called the solute's
\textit{electrostatic solvation free energy} $\Delta G^{\mathrm{solv,es}}$,
and it may be written as
\begin{equation}
\Delta G^{\mathrm{solv,es}} = \frac{1}{2} \sum_{i=1}^{n_c} q_i \varphi^{\mathrm{REAC}}(r_i).
  \end{equation}
Electrostatic solvation free energies and the bare Coulomb energies
can be used to estimate quantities such as electrostatic contributions
to protein stability~\cite{Spector00} or the binding affinity between
molecules~\cite{Caravella99}.

One may also investigate the effect of dilute ionic solutions on a
solute using either the linearized Poisson--Boltzmann (\textsc{pb}) equation
$\nabla^2 \varphi_{II}(r) = \kappa^2 \varphi_{II}(r)$ to model the
potential in the solvent, or the full nonlinear \textsc{pb}
equation~\cite{SharpHonig1990b}.  The linearized \textsc{pb} problem can also
be solved using boundary integral
equations~\cite{Yoon90,JufferETal1991,LuETal2006}.  For simplicity in
presenting the \fmm, we treat only the mixed-dielectric Poisson
problem; however, our implementation can also solve Helmholtz problems
and we are currently adapting the method for solving linearized
Poisson--Boltzmann \bem\ problems.

The first step in using \bem\ to solve Equation\eqref{eq:ASC} is to divide
the boundary $\Omega$ into $n_p$ discrete, non-overlapping pieces,
called panels or boundary elements.  Often, for complicated geometries
one approximates the boundary using easily defined boundary elements
such as planar triangles, a practice we follow here. Our solver
infrastructure supports the use of more accurate representations with
curved boundary elements, as in~\cite{LiangSubramaniam1997,AltmanBardhanWhiteTidor09}, which we
intend to explore in future work.  The second step in \bem is to
define the space of possible solutions; here, we define $n_p$
piecewise-constant functions, the $i^{\text{th}}$ of which takes a value of 1 on
element $i$ and zero elsewhere.  The third step is defining what
constraints the approximate solution should
satisfy~\cite{Atkinson97,Bardhan2009,Bardhan09_PRE}.  Galerkin
methods enforce the residual to be orthogonal to the basis set,
whereas point collocation methods enforce that the residual is exactly
zero at specified points on the boundary.

By enforcing $n_p$ constraints, one obtains a square matrix equation
\begin{equation}
  A x = B q,
  \end{equation}
where $B$ is the $n_p$-by-$n_q$ dense matrix that maps the point
charge values (the vector $q$) to the normal electric field that they
induce at the surface, and $A$ is the $n_p$-by-$n_p$ dense matrix
associated with the second-kind integral operator of Equation\eqref{eq:ASC}.
The unknown basis-function weights $x$ are found in practice by
solving the linear system using Krylov-subspace iterative methods such
as \gmres~\cite{SaadSchultz1986} and preconditioning, and instead of computing
all the entries of $A$ explicitly one uses fast-summation techniques
such as the fast-multipole method
(\fmm)~\cite{Rokhlin1983,GreengardRokhlin1987,NaborsWhite1991},
precorrected-\textsc{fft}~\cite{PhillipsWhite1997}, or \textsc{fftsvd}~\cite{AltmanETal2006} to
compute dense matrix--vector products using only linear, $\mathcal{O}(n_p)$, or
near-linear time and memory~\cite{DijkstraMattheij06}.  After
obtaining $x$, the vector of reaction potentials at the charge
locations can be computed as the matrix--vector product $\varphi^{REAC} =
C x$, where $C$ is the $n_q$-by-$n_p$ matrix resulting from
discretizing the Coulomb operator of \eqref{eq:reac}.  Therefore,
the overall electrostatic solvation free energy can be written as
$\frac{1}{2} q^T C A^{-1} B q$.

In this work, we use a Galerkin approach and the simplest possible
quadrature rule\,---\,a single point, located at the center of the
triangle\,---\,although the solver supports more sophisticated
approaches.  Representing $\sigma(r)$ using point charges is a common
approach for estimating energies, especially in quantum
chemistry~\cite{Chipman97}, and is adequate for the current
demonstration of the \gpu-accelerated \fmm using \bibee.  For planar elements and
polynomial basis functions, one may also compute some of these
integrals analytically~\cite{Hess62,Newman86} rather than by numerical
quadrature, and implementation of these approaches is underway.

\subsection{BIBEE approximation to the continuum model}\label{s:bem}

The recently developed \bibee (boundary-integral-based electrostatics
estimation) models compute approximations to a molecule's
electrostatic solvation free energy in the mixed-dielectric Poisson
model~\cite{Bardhan08_BIBEE,BardhanKnepleyAnitescu09}.  \bibee
calculations can be at least an order of magnitude faster than \bem
simulation, and the approximations reproduce numerous important
characteristics of the actual Poisson model, which make \bibee an
appealing new approach for studying electrostatic interactions within
and between large biological molecules.

A \bibee calculation approximates the solution of the boundary-integral
equation approach in the previous section.  As described earlier, the
\bem approach to calculating the electrostatic solvation free energy
requires three steps: the computation of $B q$, the normal electric
field at the boundary; solving the linear system $A x = B q$ to obtain
the induced surface charge distribution; and the computation of the
reaction potentials, $\varphi^{reac} = C x$.  \bibee approximates the
second step by replacing the electric-field operator in
\eqref{eq:ASC} with a scaled identity operator.  One therefore
obtains an approximate surface charge density $\hat{\sigma}$ rather
than the actual solution $\sigma$.  We use a scale
factor of $-1/2$, which corresponds to an extremal eigenvalue of the
electric-field operator~\cite{Atkinson97}:
\begin{equation}
\noindent \left(1 -\frac{\epsilon_I}{\epsilon_{II}}\right)
\left(\frac{\partial}{\partial n}\sum_{i=1}^{n_c}\frac{q_i}{4 \pi ||r-r_i||}
-  \frac{1}{2} \hat{\sigma}(r)
\right) 
= \hat{\sigma}(r).\label{eq:CFA}
  \end{equation}
Other useful \bibee approximations may be obtained by employing $0$ or
$+1/2$ as scale factors, which are also important eigenvalues for the
integral operator~\cite{Bardhan08_BIBEE,BardhanKnepleyAnitescu09}.

Numerical \bibee calculations can be performed using simple
modifications of \bem implementations for solving \eqref{eq:ASC}.
The \bibee/\cfa approximate solvation free energy is written as
$\frac{1}{2}q^T C D^{-1} B q$, where $C$ and $B$ are the same as in
\bem, and $D$ is a diagonal matrix~\cite{Bardhan08_BIBEE}.  \bibee's
speed advantage over \bem simulation arises because $D^{-1}$ is
trivial to apply, whereas application of $A^{-1}$, which entails the
Krylov-iterative solve, is the most computationally expensive step in
computing electrostatic solvation free energies using \bem.

Using Eq.~\eqref{eq:CFA}, the approximate electrostatic solvation free
energy for a single point charge (setting all others to zero) is equal
to the volume integral of the energy density of the Coulomb field
produced by the lone charge, where the volume of integration includes
the entire solute volume except for a spherical region associated with
the atomic charge in question (which eliminates the
singularity~\cite{Qiu97}).  This solvation free energy, because it is
associated with setting $q_i = 1$ for some $i$ and the other charges
to zero, is known as the $i^{\text{th}}$ self energy.  The use of such a volume
integral to approximate a point charge's self energy is well-known as
the Coulomb-field approximation~\cite{Qiu97}, and therefore we refer
to Eq.~\ref{eq:CFA} as the \bibee/\cfa model.

\bibee/\cfa provides new insights into the character of the \cfa,
including that the \cfa is exact for charge distributions that generate
uniform normal fields on the boundary~\cite{Bardhan08_BIBEE} and that
the \bibee/\cfa approximate electrostatic solvation free energy is an
upper bound to the actual free energy that would be obtained by
solving the Poisson model~\cite{BardhanKnepleyAnitescu09}.
Ironically, although the \cfa gives a mathematically rigorous
approximation of the Poisson problem, until recently it had been
relegated to computing parameters for non-rigorous, purely empirical
electrostatic models in the generalized-Born 
methods~\cite{StillETal1990,Qiu97}.  In fact, the \cfa was originally
introduced in volume-integral form to calculate the self-energies of point charges
\cite{Qiu97}.  Despite the unphysical basis of \textsc{gb} methods,
they have proven to be effective at approximating solvation free
energies.  As a result, later research focused on calculating more
accurate self-energies (improvements ``beyond the Coulomb-field
approximation''~\cite{Romanov04}) rather than on analysis of the \cfa.
Borgis \emph{et al.}\ achieved a significant theoretical advance with their
variational Coulomb-field approximation~\cite{Borgis03}, which allowed
for the first time the treatment of multiple charges in the \cfa,
eliminating the need for the nonphysical approximations inherent in
\textsc{gb}.  Several years later, independent analysis of the
boundary-integral equation (\ref{eq:ASC}) led to the \bibee
models~\cite{Bardhan08_BIBEE}.  This latter work was inspired by the observation of the relationship between the \cfa volume
integral and Equation \eqref{eq:ASC} for a spherical solute with a single
central charge~\cite{Ghosh98}.  \bibee/\cfa appears to be the surface
form of the earlier variational \cfa of Borgis \emph{et al.}

\subsection{Fast multipole method}\label{S:fmm}

The fast multipole method is an algorithm that accelerates the computations required in $N$-body problems,  which are expressed as a sum of the form
\begin{equation}\label{eq:interaction}
  f(y_j) = \sum_{i=1}^N c_i \, \uKern(y_j, x_i).
\end{equation}
\noindent Here, $f(y_j)$ represents a field value evaluated at a point
$y_j$, where the field is generated by the influence of sources
located at the set of centers $\{x_i\}$. The evaluation of the field
at the centers themselves corresponds to the well-known $N$-body
problem.  Thus, $\{x_i\}$ is the set of source points with weights
given by $c_i$, $\{y_j\}$ the set of evaluation points, and
$\uKern(y,x)$ is the kernel that governs the interactions between
evaluation and source points. Obtaining the field $f$ at all the
evaluation points requires in principle $\mathcal{O}(N^2)$ operations,
if both sets of points have $N$ elements each.  Fast summation algorithms aim at
obtaining $f$ approximately with a reduced operation count, ideally
$\mathcal{O}(N)$. In our application, we use two types of
fast summation: in the first, we replace the source points with a
discretization of the induced charge on the molecular surface
$\sigma$, evaluate the field at the same surface points, and use as
the kernel the boundary integral operator in Equation \ref{eq:ASC}.
In the second type of fast summation, the source points are the point
charges in the molecule and the field is evaluated at points on the
surface~$\Omega$.

There are two main aspects to the \fmm algorithm: one is related to the spatial hierarchy and tree data structure; the other, to the mathematical machinery needed to perform approximations which reduce the number of operations in exchange for accuracy. The spatial hierarchy refers to a successive sub-division of the three-dimensional spatial domain and construction of an associated oct-tree: the domain is initially divided in eight cells, which are then divided again, and so on, until the finest level of refinement or ``leaf level''. The set of cells created by each sub-division is associated to the nodes at a given level of a tree structure, such that near- and far-domains to each cell can be found by operations on the tree. The source points belonging to cells in the far-domain are then considered as a cluster, with their influence represented by a series expansion in the process of evaluation. This representation of many sources collectively by series expansions allows the reduction of operations in the \fmm to \bigON, with controllable accuracy.

\begin{figure*}
	\centering
	{\includegraphics[width=0.95\textwidth]{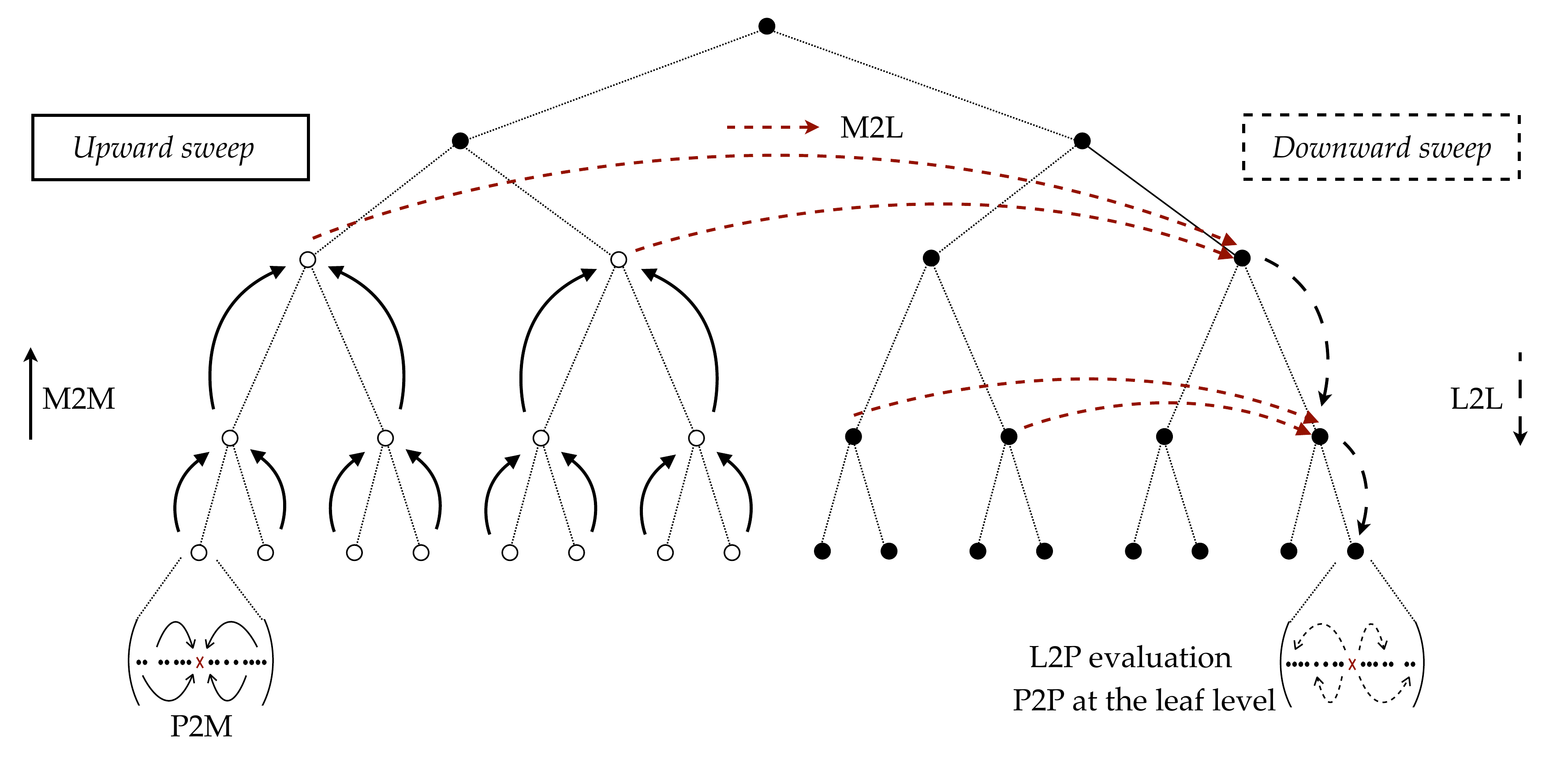}}
	\caption{On a tree diagram, we can illustrate the organization of the algorithmic stages for the \fmm: \emph{upward sweep}, \emph{downward sweep}, and \emph{evaluation step}.  The upward sweep combines the \PM and \MM operations, downward sweep the \ML and \LL, and the evaluation combines \LP with a direct calculation in the near domain, \PP.}
	\label{fig:UpDown-sweeps}
\end{figure*}

One can illustrate the phases of the algorithm using a diagram of the tree associated to the spatial division, thereby directly relating the algorithm to the data structure used by the \fmm; see  Figure \ref{fig:UpDown-sweeps}.  First, we need to introduce the terminology for the approximation of the kernel action at long and short distances, as follows.

\begin{itemize}
\item[$\triangleright$] \textit{Multipole Expansion (\ME)}: a series expansion truncated after $p$ terms that represents the influence of a   cluster of source points, and is valid at distances large with respect to the cluster radius.
\item[$\triangleright$] \textit{Local Expansion (\LE)}: a truncated series expansion, valid only inside a sub-domain, that is used to efficiently evaluate a group of \ME s locally to a cluster of evaluation points.
\end{itemize}

The computation of the action of the kernel using \fmm\ proceeds in stages: an \emph{upward sweep}, \emph{downward sweep}, and \emph{evaluation stage}. In the \emph{upward sweep}, the objective is to build the \ME s for each node of the tree. The \ME s are built first at the tree leaves (\PM operation) and then translated to the center of the parent cells (\MM translation).  This process is illustrated in Figure~\ref{fig:UpDown-sweeps} by the black arrows pointing up from the nodes on the left side of the figure. Notice that at each level above the leaves, \ME s are computed by shifting and then combining the \ME s of the child cells, which results in a reduction  by a factor $8$ of the number of expansions that needs to be calculated. In the \emph{downward sweep} of the tree, the \ME s are first transformed into \LE s for all the boxes in the \emph{interaction list}\,---a process represented by the dashed red-colored arrows in Figure \ref{fig:UpDown-sweeps}, and denoted by \ML. For a given cell, the interaction list corresponds to the cells whose parents are neighbors of the given cell's parent, and yet not directly adjacent to the given cell. Each \LE\ is then translated to the centers of all child cells (\LL operation), and combined with the transformed \ME s from the child level to obtain the complete far-domain influence for each box. This  process is represented by the dashed arrows going down the right side of the tree in Figure \ref{fig:UpDown-sweeps}. At the end of the \emph{downward  sweep}, each cell will have an \LE\ that represents its complete far field. Finally, at the \emph{evaluation stage}, the field is evaluated for every point contained in a cell by adding the near-field and far-field contributions: the near field contribution is obtained from directly computing the interactions between all the points in the adjacent cells, and the far-field contribution comes from evaluating the \LE\ of the cell at each point location.

\section{Multi-\gpu\ fast multipole method} 

\subsection{Brief overview of the distributed \fmm}

The distributed-memory parallelization of the \fmm involves two main phases: \emph{(i)} the partitioning of the tree among MPI processes; and, \emph{(ii)} the communication of needed data to perform the \fmm interactions. In the first phase, the standard and most commonly used technique is to equally distribute the Morton-indexed boxes at the leaf level \cite{WarrenSalmon1993}. This has been the approach used here.

The second phase involves the communication that occurs between partitioned domains, due to interacting cells residing in different processes. Two cells interact when they are present in each other's neighbor list or interaction list (for the far-field), as defined in the previous section. In the case of the neighbors, communication only occurs among processes holding geometrically adjacent points, and data consist of position and weights of the points or particles. For the far-field, the data that needs to be communicated consists of \ME\ coefficients of the cells in the interaction list, at every level of the tree. We overlap these communications with local computations, which results in effective hiding of communication time thereby enhancing scaling (as seen in the results section).  Moreover, we have achieved linear scaling in memory requirements, which allows our large-scale computations of biomolecular electrostatics.

\subsection{Synopsis of \gpu\ hardware utilization}

\begin{figure}
	\centering
	{\includegraphics[width=0.8\columnwidth]{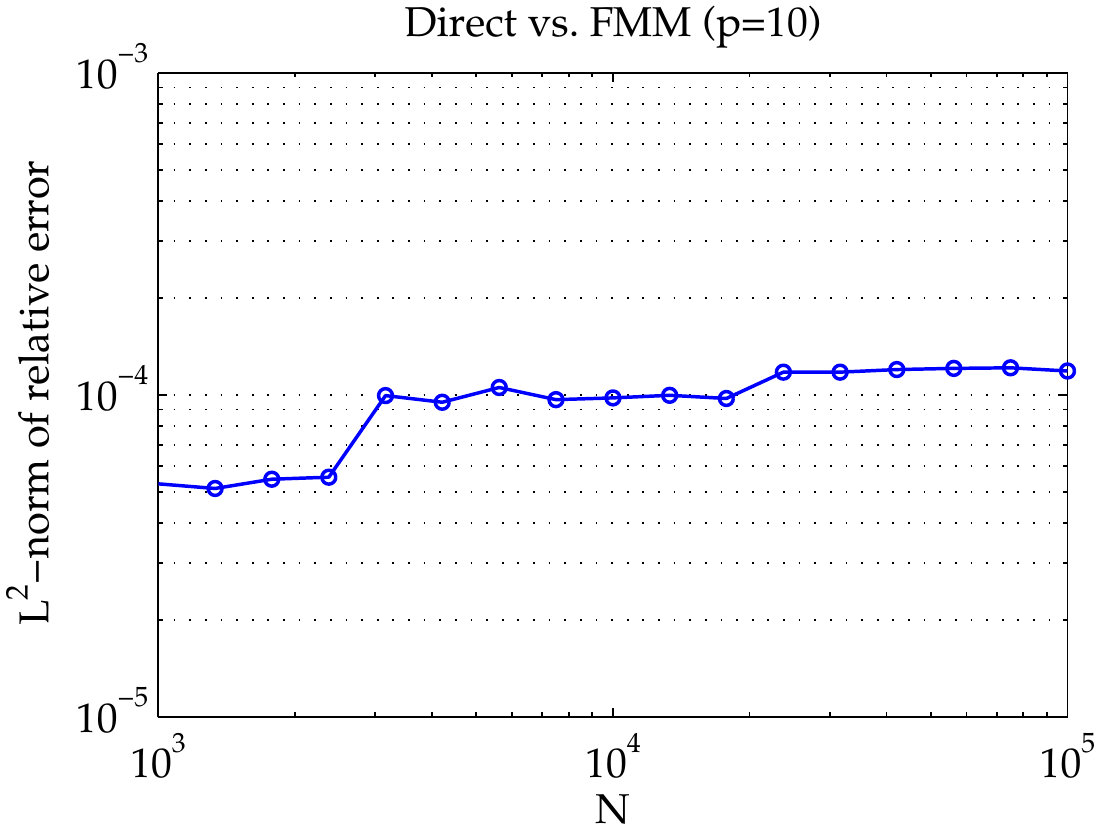}}
	\caption{$L^{2}$-norm of the measured normalized error for the \fmm evaluation \emph{versus} direct evaluation on the \gpu.}
	\label{fig:errorFMM}
\end{figure}

All the computational kernels of the \fmm depicted in Figure \ref{fig:UpDown-sweeps} were reformulated to utilize the \gpu\ architecture with \cuda \cite{cuda-guide}, using  single precision. The \gpu\ kernels were verified against \cpu\ calculations, and full \fmm evaluations were also validated with respect to direct (all-pairs) evaluation. Figure \ref{fig:errorFMM} shows the measured error ($L^{2}$-norm of the relative error) for a set of calculations with various numbers of points, comparing the \fmm on \gpu\ with direct evaluation.

One particular feature of the \gpu\ execution that we've incorporated in the code is an effective buffering of input and output data to each kernel, such that a kernel is always executed on sufficiently large data sets. In other words, a given kernel, \emph{e.g.}, \ML, is not executed for each of the interactions, but for a group of them. At the same time, data is made contiguous in memory before the kernel call and output memory is also made contiguous in the \gpu\ buffer; these are essential measures to attain good performance. On the \gpu, the memory for evaluation points is padded to match the size of a thread block. For example, if the number of multipole coefficients is 55 and the thread block size is 64, the code will insert zero padding for indices 56--64. This coalesced write to evaluation points was found to be much more efficient than using a compressed buffer.

\section{Computational results}

\subsection{Hardware}
\begin{figure*}
\begin{center}
\includegraphics[width=0.7\textwidth]{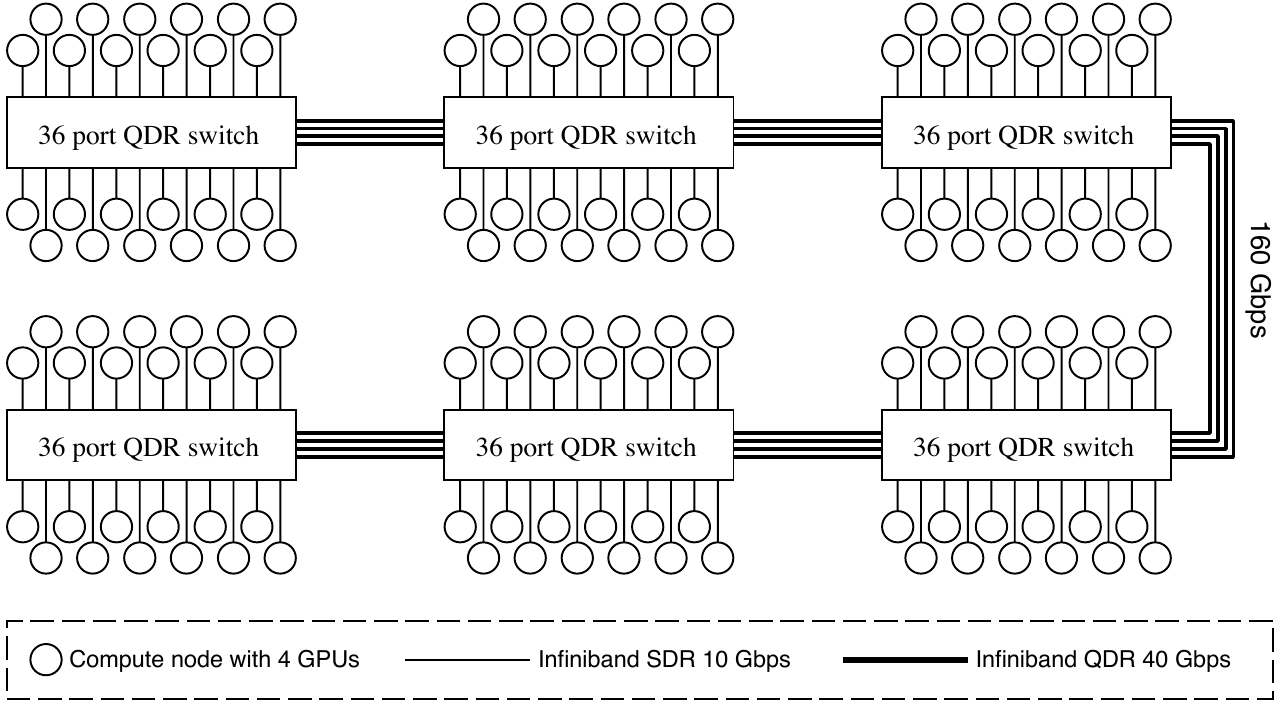}
\end{center}
\caption{Infiniband network configuration of the \emph{Degima} cluster in Nagasaki.}
\label{fig:network}
\end{figure*}

The calculations were performed using the \textit{Degima} cluster at the Nagasaki Advanced Computing Center, which presently consists of 288 \nvidia \textsc{gtx} 295 cards, each with two \gpu s. There are two cards per host node, amounting to 144 \cpu s and 576 \gpu s---however, in this work we have only used 128 nodes out of the total 144, as some of the nodes remain in experimental use.  Figure \ref{fig:network} shows the configuration of the interconnect between the nodes. There are 36 nodes connected to each of the 6 \textsc{qdr} switches. The bandwidth of \textsc{sdr} is 10 Gbps and for the \textsc{qdr} it is 40 Gbps. With 4 \textsc{qdr} networks, a total bandwidth of 160 Gbps is achieved between the switches. Each circle in the Figure represents one compute node equipped with 1 \cpu\ and 4 \gpu s. 

\subsection{Overview of computational experiments}

The calculations we report here demonstrate the speed and scalability
of our \fmm algorithm on realistic biomolecular problems, without any attempt at
this point of offering biological insights into our chosen examples.
Much more detailed simulation and analysis are required to obtain
meaningful understanding of these complex systems.  Furthermore, it
bears repeating that techniques such as explicit-solvent molecular
dynamics offer much more detail and are to be preferred in some
circumstances, but are not always practical for some types of
problems.

We first present the scalability of the \fmm on a large number of \gpu s to demonstrate the computational power of the present method. Subsequently, we study the convergence and accuracy of the \fmm-based \bem approximation \bibee
using as a model problem a clinically relevant protein
(implicated in cancer) bound to a small-molecule inhibitor.  We then
calculate the electrostatics of several multi-million atom systems by
creating arrays of molecules of the protein lysozyme.  This example is
inspired by the pioneering work of McGuffee and
Elcock~\cite{McGuffee06}, who were among the first to conduct
atomistic-level simulations of concentrated protein solutions.  This
kind of large-scale, computationally demanding study is important
because proteins in biological systems, especially inside cells,
operate in crowded environments that can strongly influence protein
behavior~\cite{Minton00}.  Computational expense has long been a
severe constraint on scientists' ability to study such systems
theoretically.

Most investigations use implicit-solvent models in conjunction with
Brownian dynamics (see, for example, \cite{Gabdoulline97}) to maximize
the time scales that can be simulated at reasonable computational
cost.  Still, however, most studies have of necessity employed reduced
models of the proteins (\emph{e.g.}, spheres) despite evidence that
protein shape plays a significant role~\cite{Neal95,Chang00}.
Atomistic-level treatments have also been forced to use some
approximations to the physics (approximating the electrostatics) to
make the computations feasible~\cite{Elcock04}.  To our knowledge, our
ability to compute these interactions rigorously in seconds is
unprecedented, and may enable more accurate studies than have been
possible previously.

The Appendix provides details regarding the preparation of the protein
structures as well as the surface discretizations.  All calculations
were performed with $\epsilon_I = 4$ and $\epsilon_{II} = 80$.

\subsection{Scalability of the \fmm on a \gpu\ cluster} 

In this section, we present and discuss results from a scalability study of the \fmm on many \gpu s. We investigate scalability under the more severe \emph{strong scaling} scenario, where the amount of computational work in the experiments remains constant as the number of computing devices (here, \gpu\ chips) is increased from 1 to 512. In addition to parallel efficiency, we analyze the breakdown of execution time by computational kernels, and also by \cuda states. The results show excellent parallel efficiency up to 256 \gpu s, and subsequent decline as serial fraction and communication overheads start to dominate. Note that the 256  \gpu\ chips offer a total of 61,440 \cuda cores; 
the case on 512 \gpu s is in actual fact running several million simultaneous threads. 

It must be emphasized that achieving high parallel efficiency on a cluster of \gpu s is generally much more difficult than on a cluster of \cpu s. As pointed out above, each \gpu\ chip is already a parallel device, in this case (\textsc{gt200}b \gpu) offering 240 multi-threaded cores. This results in a high compute capability which makes it very difficult to hide the time required for communications. Moreover, the current technology does not provide an avenue for direct communication between \gpu\ cards, and all data must be transfered to and from the host \cpu\ first; communication between nodes then occurs via standard MPI calls. Despite these serious challenges, we can report excellent parallel efficiency, as already mentioned.

\paragraph{Experimental setup} The scalability study is based on a fixed-size problem with $N=10^8$ points placed randomly in a cubic volume. The order of the multipole expansions was set to $p = 10$, which results in 4 significant digits of accuracy, and spherical harmonic rotations are performed before each translation at $\mathcal{O}(p^3)$ cost. One \fmm evaluation is performed (equivalent to one matrix-vector product, or one \textsc{bibee} evaluation, or the cost of one iteration in a \bem\ solution) for each case, with one MPI process running per \gpu. We run one MPI process per cluster node up to 128 processes, and then two and four processes per node, respectively, in the 256- and 512-\gpu\ cases (recall that each node has two dual-\gpu\ cards).

\begin{figure}
	\centering
	{\includegraphics[width=0.8\columnwidth]{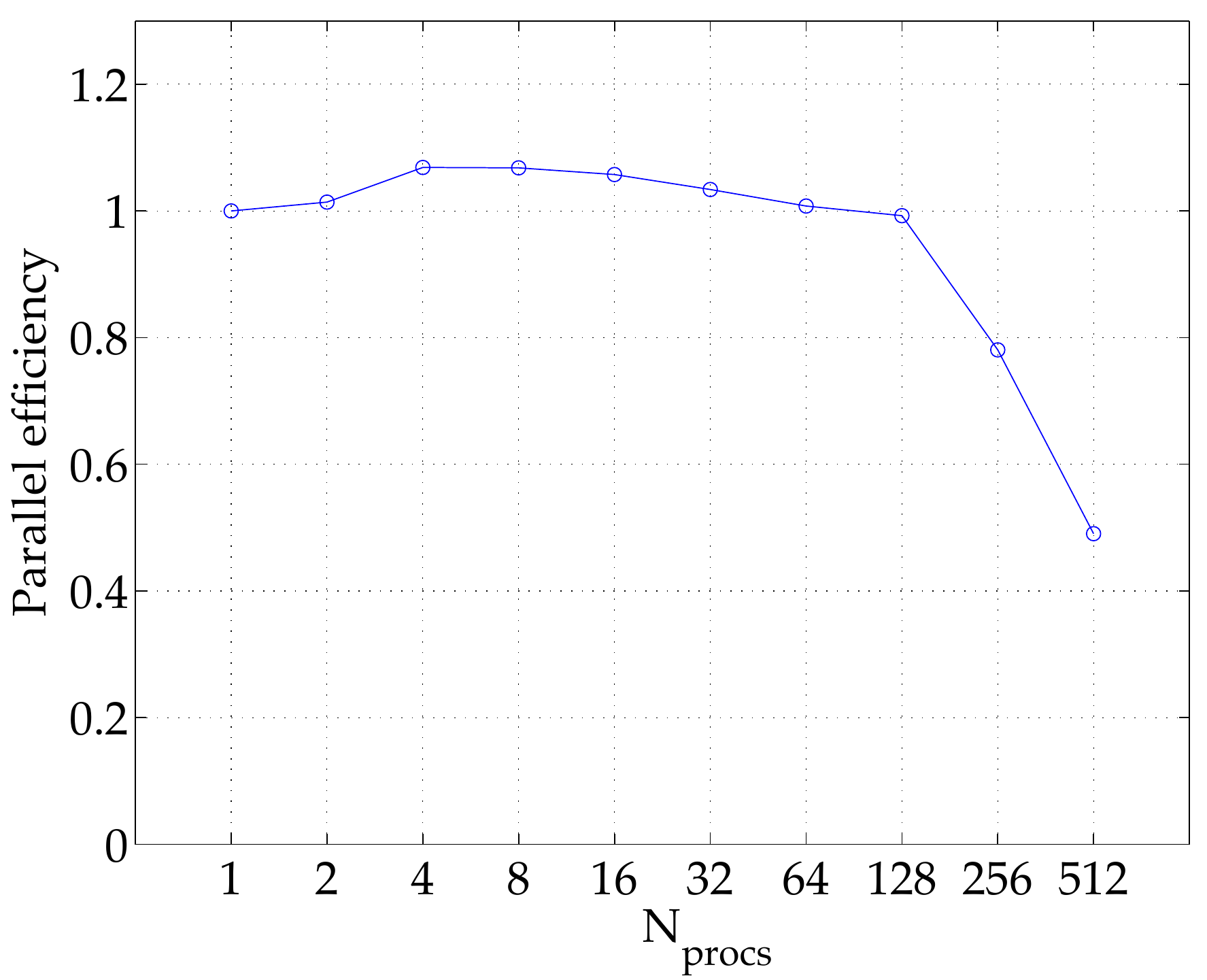}}
	\caption{Parallel efficiency in strong scaling of the \fmm on multi-\gpu s; $N=10^8$.}
	\label{fig:strong_scaling}
\end{figure}

\begin{figure}[t]
	\centering
	{\includegraphics[width=0.98\columnwidth]{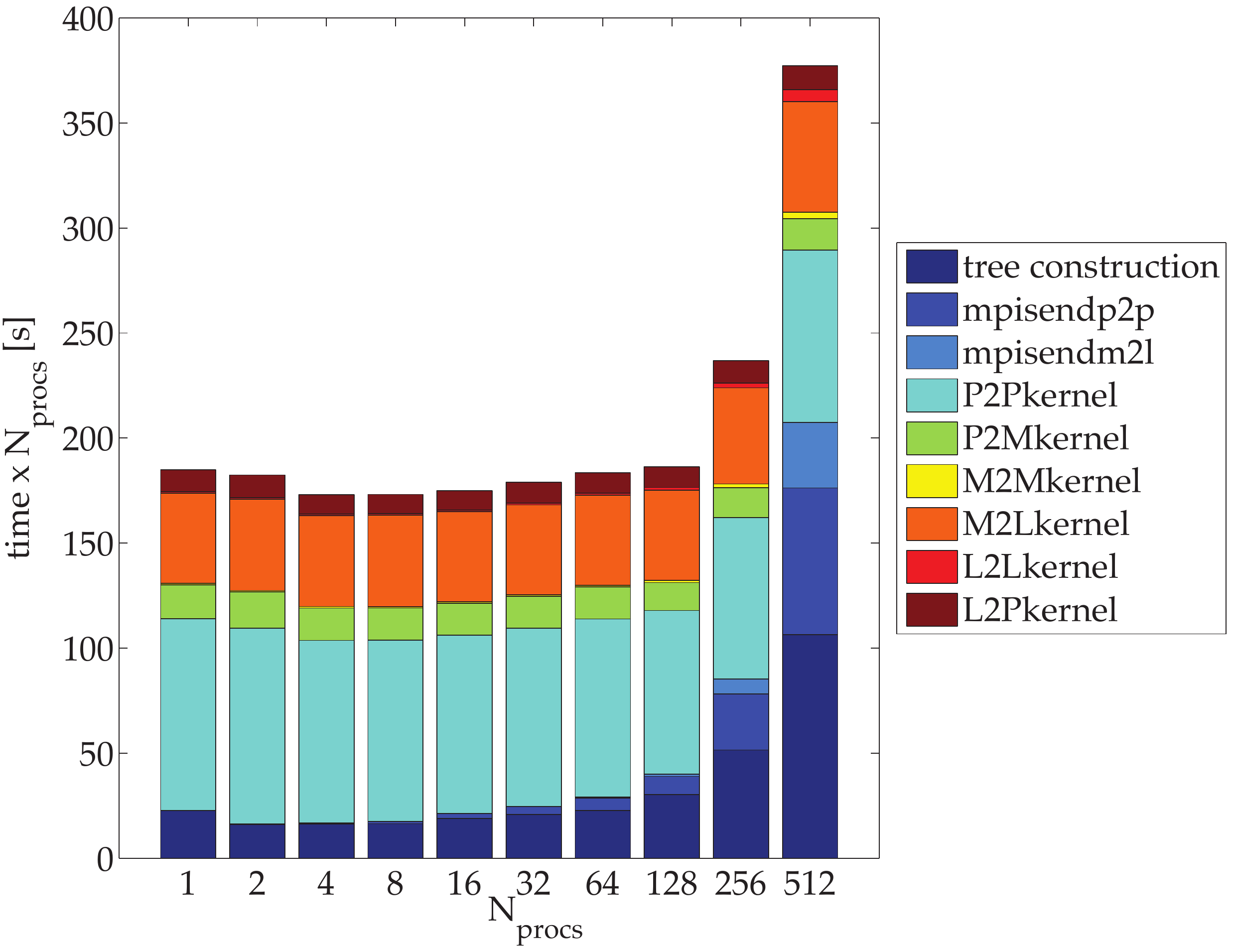}}
	\caption{Breakdown of the \fmm calculation time corresponding to the runs in Figure \ref{fig:strong_scaling}, with $N=10^8$.}
	\label{fig:FMMbreakdown}
\end{figure}

\paragraph{Results} As shown in Figure \ref{fig:strong_scaling}, parallel efficiency remains perfect up to 128 \gpu s, is 78\% for 256 \gpu s, and 48\% for 512 \gpu s. We observe ``super-linear'' scaling between 4 and 32 \gpu s, which we ascribe to better cache usage in the tree construction phase of the calculation, done on the \cpu. This is evidenced by the breakdown of time shown in Figure \ref{fig:FMMbreakdown}, where the tree construction takes less per-process time when going from 1 to 2 and 4 \gpu s.

Figure \ref{fig:FMMbreakdown} is a bar plot of the runtime multiplied by the number of \gpu s (number of MPI processes), where each bar is also color-coded to indicate the time spent in each computational kernel, including tree construction (done on \cpu) and MPI communications. Equal bar heights would indicate perfect strong scaling, so once again we observe super-linear scaling for 4--32 processes and efficiency decline for more than 128 \gpu s. The breakdown helps to make these observations: tree construction on the \cpu\ is overburdened on one process, possibly due to cache misuse; as expected, the \PP and \ML kernels take the largest fractions of runtime; the fraction of time for MPI communications is minor, up until 256 processes; and, tree construction and communications become significant at 512 processes.

\begin{figure}
	\centering
	{\includegraphics[width=0.98\columnwidth]{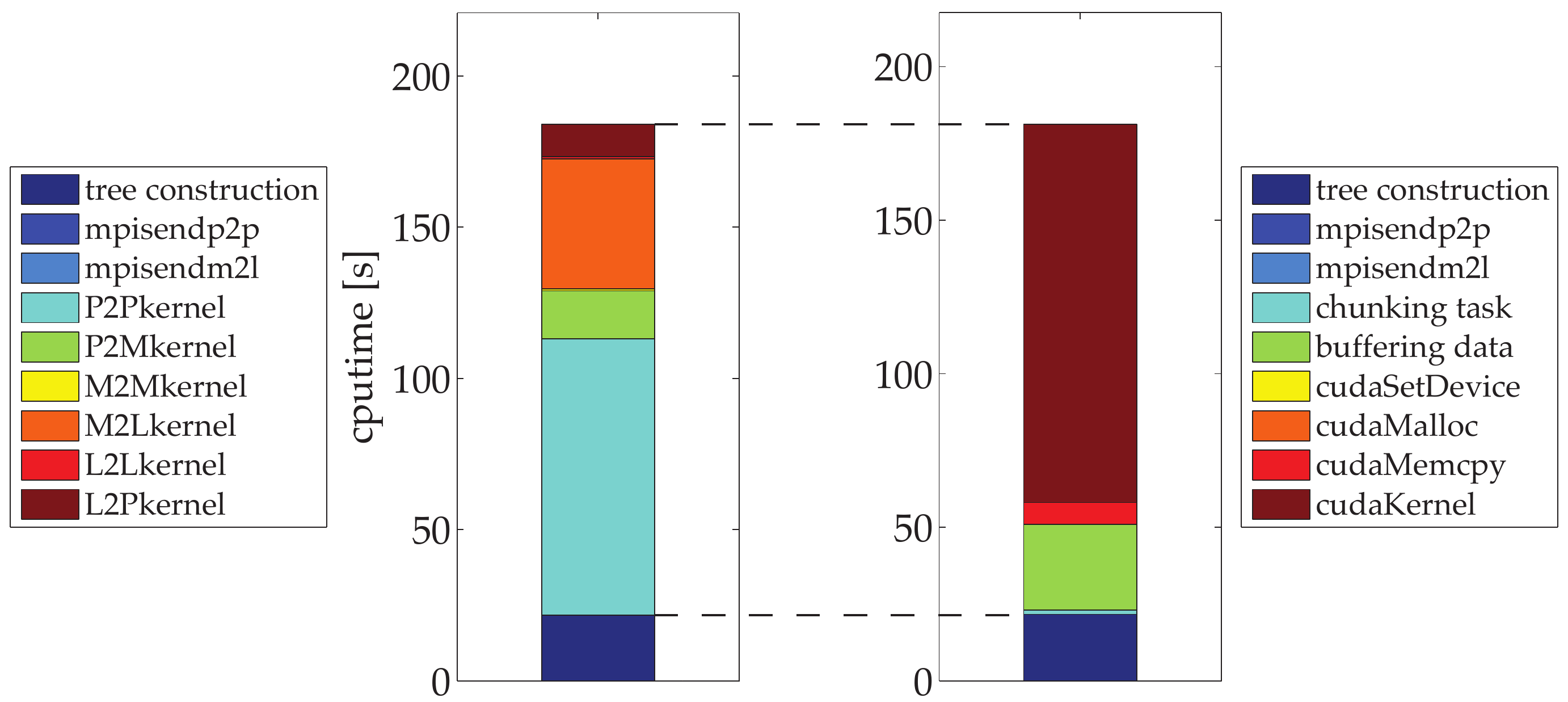}}
	\caption{Breakdown of the \fmm calculation time corresponding to the run with $N_{procs}=1$ in Figure \ref{fig:FMMbreakdown}, with $N=10^8$.}
	\label{fig:GPUbreakdown}
\end{figure}

The tree construction consists of all the preprocessing required to execute the \fmm kernels,  including the Morton indexing and bookkeeping of the interaction list, taking as input the set of points already assigned to corresponding processes. The MPI communication time, on the other hand, consists of the time required to update the data of points in the local essential tree (``mpisendp2p'') and the time required to update the multipole expansions (``mpisendm2l''). In sum, the height of each bar in Figure \ref{fig:FMMbreakdown} is the total wall-clock time it takes to perform one complete evaluation using the \fmm (multiplied by the number of processes).

The breakdown of the runtime can be also done by the phases of \cuda execution, as shown in Figure  \ref{fig:GPUbreakdown}, rather than computational kernels. Of course, the communication time and tree construction on the \cpu\ do not change in this version; only the fraction of time between the dashed lines in the Figure is analyzed differently. We have one large fraction of time (maroon color) labeled ``cudaKernel''; this is the cumulative time spent in \cuda execution state---expressed in the \cuda code by the triple triangular brackets---for all the 6 computational kernels (shown in the left bar) combined. The \fmm evaluation spends 70\% of the total execution time in this state. Next in importance is the fraction spent ``buffering data'' (in green), which refers to the reordering of data that is performed to improve coarse-grained data locality when executing the \gpu\ kernels. The required data is streamlined into a buffer in the exact order that it will be accessed inside the \gpu. Next, but already a minor fraction of the total time, is ``cudaMemcpy'' (red), a label which is given to the total time of all data movement between the host memory and the device memory. This is the most important observation to make from the breakdown of timings in this manner: the total time of all data movement between \gpu\ and \cpu\ is a small fraction of the total time, whereas a large fraction of time is spent in the \cuda execution state. This is a favorable situation in terms of \gpu\ performance, and shows that the \fmm, unlike many algorithms, can afford to move data to/from the \gpu\ due to a high compute/data ratio.

\paragraph{Discussion} We have shown strong scaling results, demonstrating excellent parallel efficiency of the \fmm up to 256 \gpu s and acceptable efficiency at 512 \gpu s. These results were obtained in an experimental cluster using gaming \gpu\ cards\footnote{The Degima cluster enabled some of the authors to be awarded the Gordon Bell prize in the price/performance category in 2009.}. The configuration of the cluster has two dual-chip \gpu\ cards per node, yet we used one \gpu\ per node up to 128 nodes to allow for the whole bandwidth to be available to each MPI process. In the 256- and 512-\gpu\ cases, the bandwidth must be shared within the node, and so we applied a strategy to alleviate possible constraints due to this. A hierarchical all-to-all communication was implemented using \texttt{MPI\_Comm\_split}, dividing the MPI communicator in $4$.  Effectively we are splitting the communicator into an inter-node and an intra-node communicator. Thus, we first perform an intra-node \texttt{MPI\_Alltoallv}, then an inter-node \texttt{MPI\_Alltoallv}. We found that this strategy was able to speed-up the communication nearly $4$ times at $512$ processes; and these are the final results presented in Figures \ref{fig:strong_scaling} and \ref{fig:FMMbreakdown}.

\subsection{Protein--inhibitor binding calculations}

Understanding the interactions between inhibitors and their proteins
can help lead to the design of more potent drugs with fewer side
effects, and computational modeling can be a valuable approach to
study inhibitor--protein binding~\cite{Jorgensen04}.  The first model
problem that we will use here is a protein known as cyclin-dependent kinase 2 (\cdk) and a
small-molecule inhibitor.  Cyclin-dependent kinases are involved in
the control of the cell cycle and implicated in the growth of
cancers~\cite{Anderson03}, and it is thought that developing
tight-binding inhibitors of \cdk proteins may be a valuable therapeutic
approach to treating some types of tumors.  The atomic structure of a
\cdk 2 protein bound to a novel small-molecule inhibitor was solved
using X-ray crystallography and deposited in the Protein Data
Bank~\cite{Berman00,Anderson03} (see Appendix).
Figure~\ref{fig:1oit-cartoon} shows the drug in the binding site.

\begin{figure}
	\centering
	{\includegraphics[width=0.75\columnwidth]{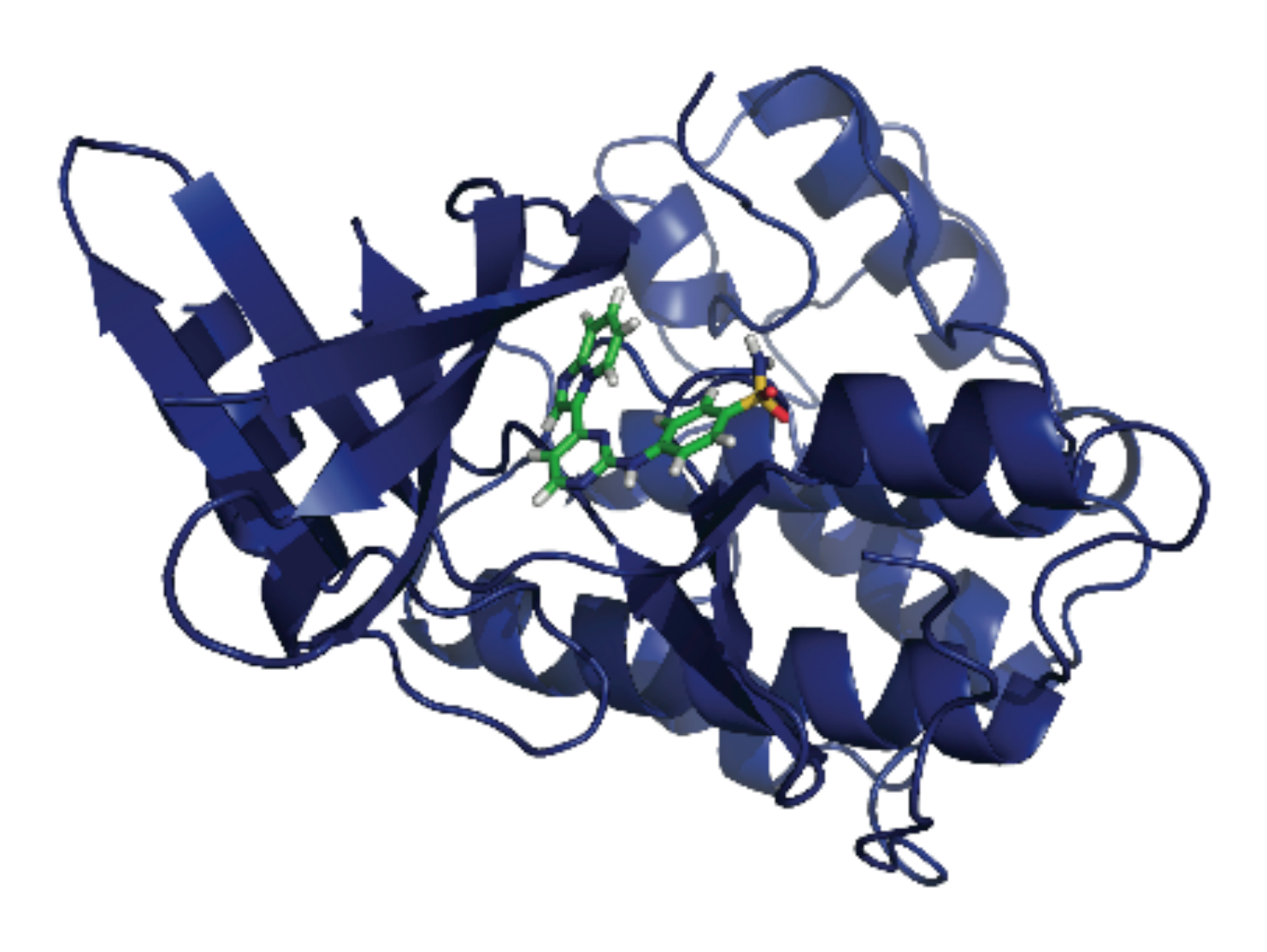}}
	\caption{Rendering of \cdk 2 bound to small molecule inhibitor, using
     atomic coordinates from PDB accession 1OIT~\cite{Anderson03} and
     the software PyMOL.}
	\label{fig:1oit-cartoon}
\end{figure}

As described in \S\ref{s:bem}, the integral equation \eqref{eq:ASC} is
approximated using \bibee/\cfa to estimate the contribution to binding
affinity that is due to the polarization of the solvent around the
protein and drug.  We use a widely accepted, but somewhat simplistic,
approach to estimating binding affinity, in which the protein and
inhibitor bind rigidly in the conformations shown in the crystal
structure.  It has been shown that more realistic models, which
account for conformational changes on binding, require highly accurate
electrostatic simulations for convergence~\cite{AltmanBardhanWhiteTidor09}.  Thus,
our solver with its \gpu\ capability may be a valuable tool to improve drug modeling by
reducing the computational cost of accounting for flexibility.  For
these types of calculations, it would be ideal to be able to
demonstrate that the computed energies are correct to within 0.1
kcal/mol, which is roughly the accuracy of many experimental binding
assays.

\begin{figure*}
	\centering
  \subfloat[][The inhibitor]{\includegraphics[width=0.7\columnwidth]{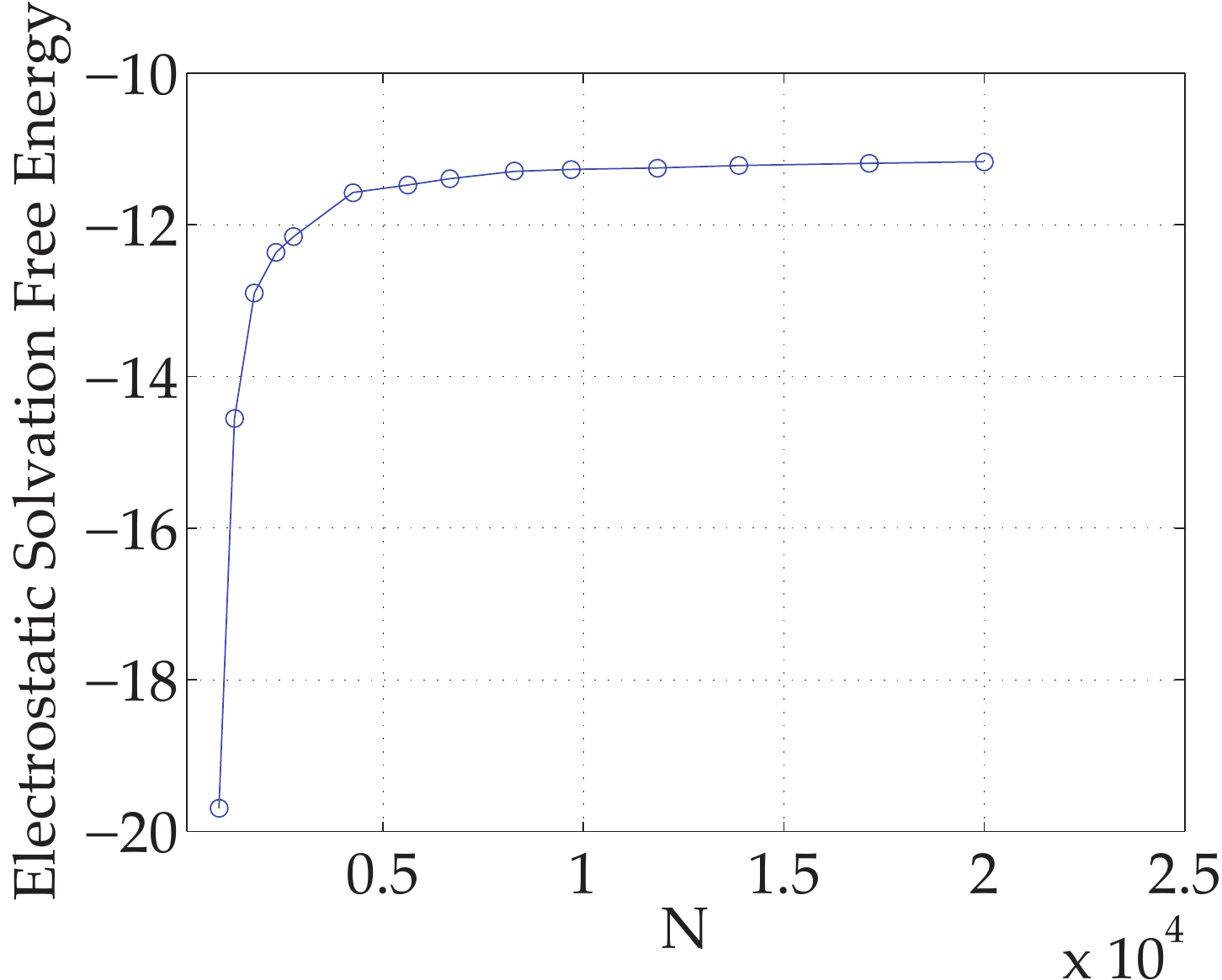}\label{fig:1oit-inh}}\qquad
  \subfloat[][\cdk 2 protein] {\includegraphics[width=0.7\columnwidth]{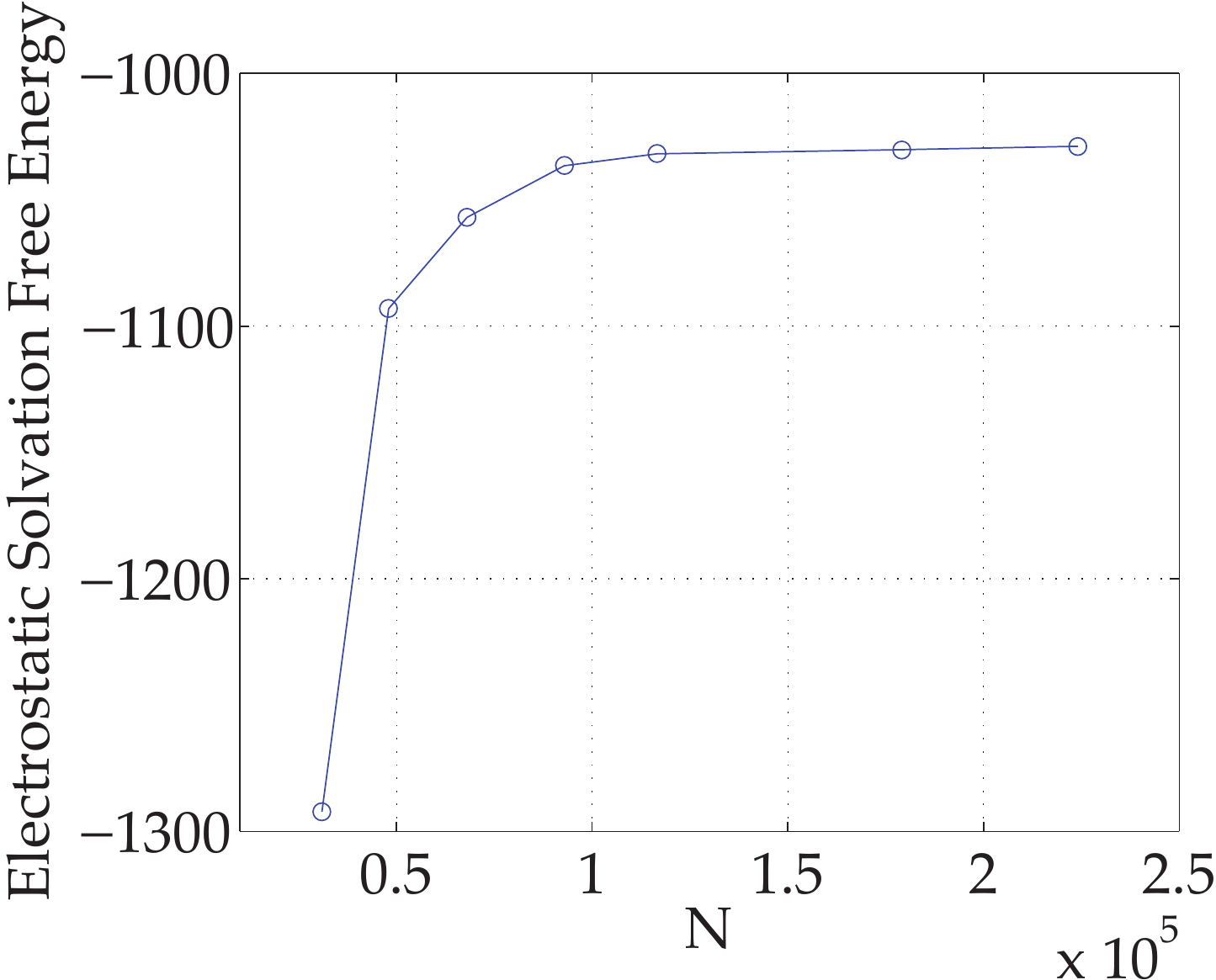}\label{fig:1oit-cdk}}\\
  \subfloat[][Inhibitor--protein complex]{\includegraphics[width=0.7\columnwidth]{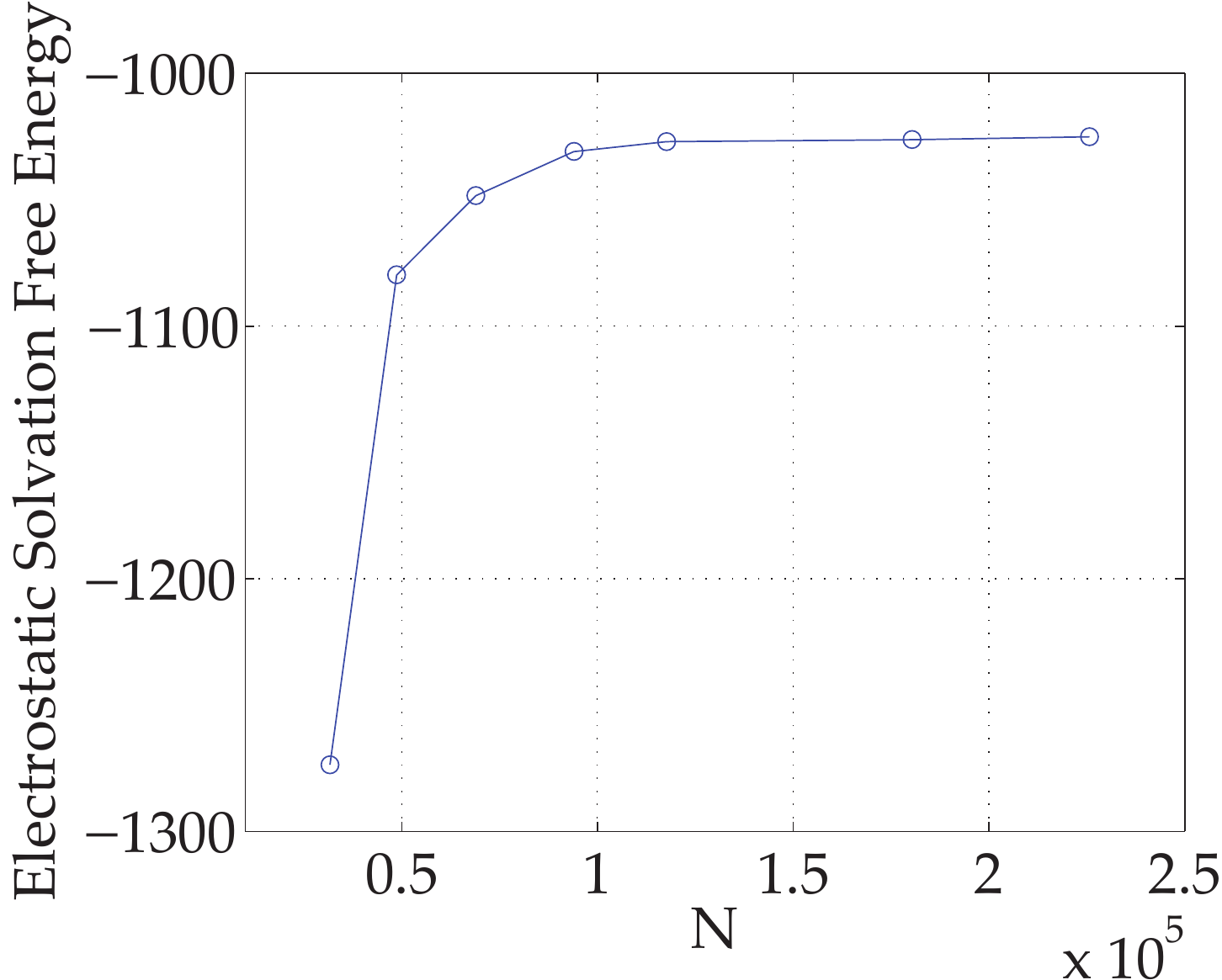}\label{fig:1oit-complex}}\qquad
  \subfloat[][Solvation free energy]{\includegraphics[width=0.7\columnwidth]{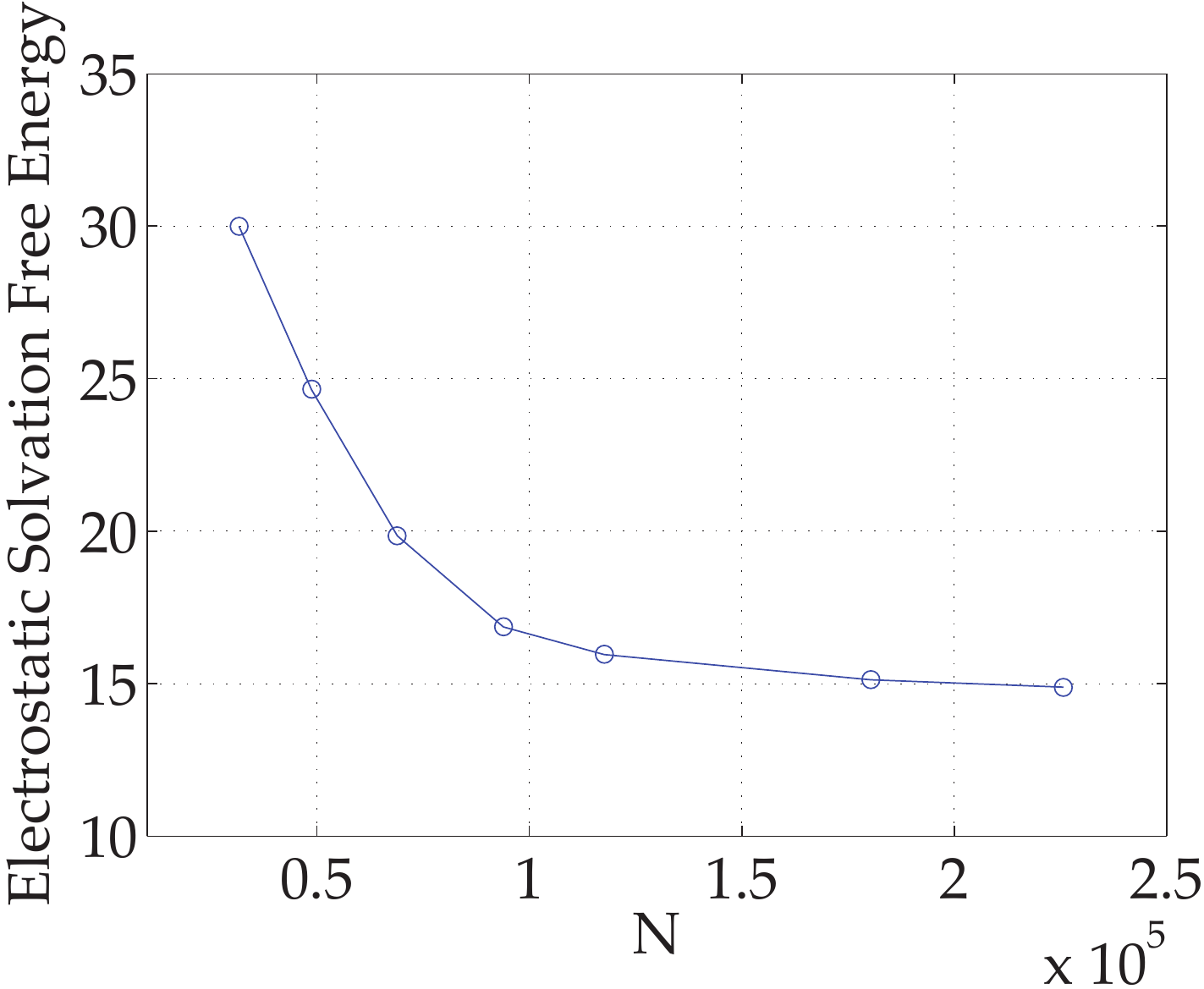}\label{fig:1oit-energy}}
	\caption{\subref{fig:1oit-inh}--\subref{fig:1oit-complex}: Convergence of calculated free energies as
     functions of the number of vertices $N$ on the dielectric
     boundaries for the components indicated. \subref{fig:1oit-energy}: Estimated
     electrostatic solvation free energy contribution to the
     inhibitor--protein binding affinity; the number of vertices $N$
     is that in the inhibitor--protein complex mesh.  All energies are in kcal/mol.}
	\label{fig:1oit-convergence}
\end{figure*}

Figures~\ref{fig:1oit-convergence}\subref{fig:1oit-inh}, \subref{fig:1oit-cdk}, and \subref{fig:1oit-complex} contain plots of the
computed electrostatic solvation free energies for the unbound
inhibitor, unbound protein, and inhibitor--protein complex, as
functions of the number of vertices $N$ on the discretized surface.  Figure~\ref{fig:1oit-convergence}\subref{fig:1oit-complex}
plots the solvation free energy of the complex minus the solvation
free energies of the unbound species, \emph{i.e.},
\begin{equation}
\Delta G^{\mathrm{solv,es}}_{\mathrm{bind}} =
\Delta G^{\mathrm{solv,es}}_{\mathrm{complex}} -
\Delta G^{\mathrm{solv,es}}_{\mathrm{protein}} -
\Delta G^{\mathrm{solv,es}}_{\mathrm{inhibitor}} \label{eq:binding-free-energy-estimate}
\end{equation}

\begin{figure}
	\centering
	{\includegraphics[width=0.9\columnwidth]{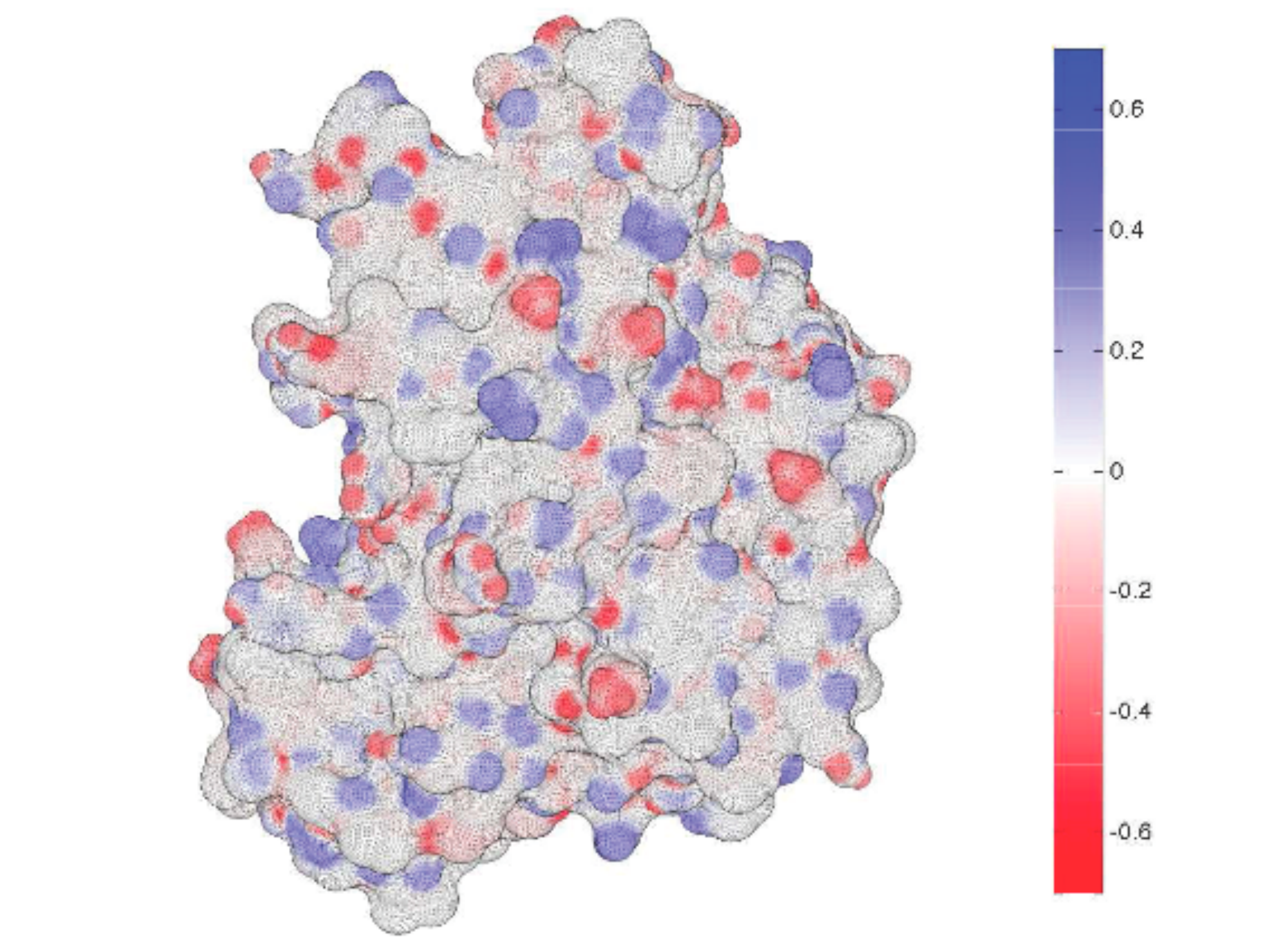}}
	\caption{The induced charge distribution at the dielectric boundary
     for the inhibitor--protein complex.  Results are in electrons per
     square Angstrom.}
	\label{fig:1oit-surface}
\end{figure}

Meshing becomes problematic for the protein and inhibitor--protein
complex beyond a vertex density of 10/square Angstrom, corresponding to the last data point on Figure \ref{fig:1oit-convergence}\subref{fig:1oit-energy}. 
Finer discretizations of the
inhibitor surface can be generated and we have simulated meshes up to
36 vertices/square Angstrom (at which point the surface is discretized
into 19,998 elements), with convergence approaching the desired
0.1~kcal/mol level.  This is
consistent with earlier work assessing the accuracy of planar elements
versus curved elements~\cite{BardhanETal2007}.  However, we note that here
we have used a point-charge approximation of the molecular charge
distribution, rather than Galerkin or collocation \bem, and have also
used the \bibee approximation rather than full \bem calculation.
Figure~\ref{fig:1oit-surface} shows a plot of the calculated surface
charge density for the complex, in electrons per square Angstrom.

\subsection{A model calculation for protein solutions}

To demonstrate the capability of the method and software on large problems, we use collections of randomly oriented lysozyme molecules arranged on a regular
Cartesian grid, mimicking the Brownian dynamics calculations performed
by McGuffee and Elcock~\cite{McGuffee06} at each time step.  In
practice, calculations on such length scales may benefit from
alternative theories that are very efficient for these specialized problems, for
instance the recent approaches of Onufriev \textit{et
  al.}~\cite{FenleyETal2008,AnandakrishnanETal2010}, but for the present the size
of the problem showcases the implementation's performance and parallel
scaling.

\begin{figure}
	\centering
	{\includegraphics[width=0.95\columnwidth]{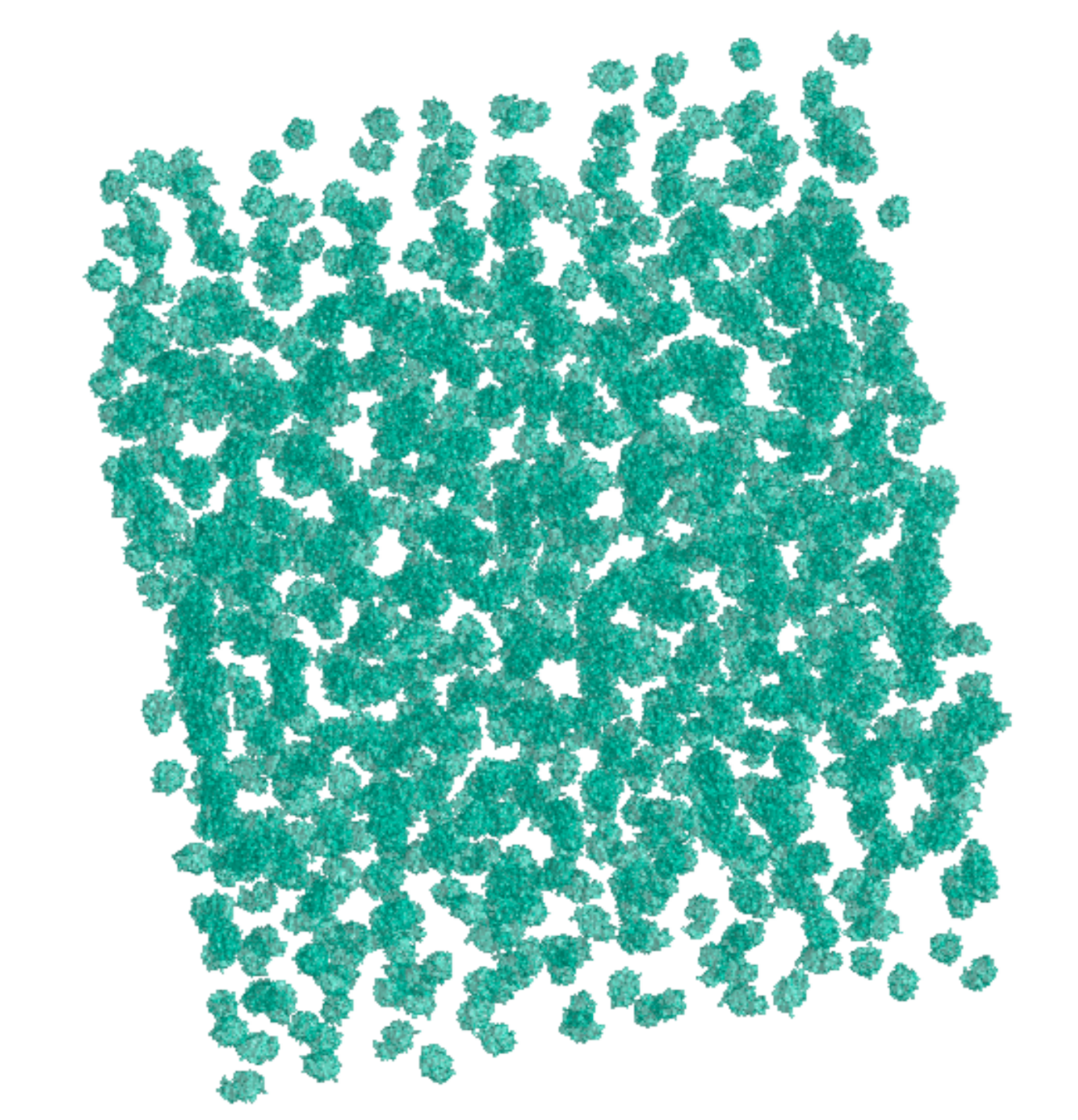}}
	\caption{Rendering of the discretized surfaces of a collection of
     1000 randomly oriented lysozyme molecules quasi-scattered from a regular
     Cartesian grid.}
	\label{fig:lysozyme-cartoon}
\end{figure}

\begin{figure}
	\centering
	{\includegraphics[width=0.9\columnwidth]{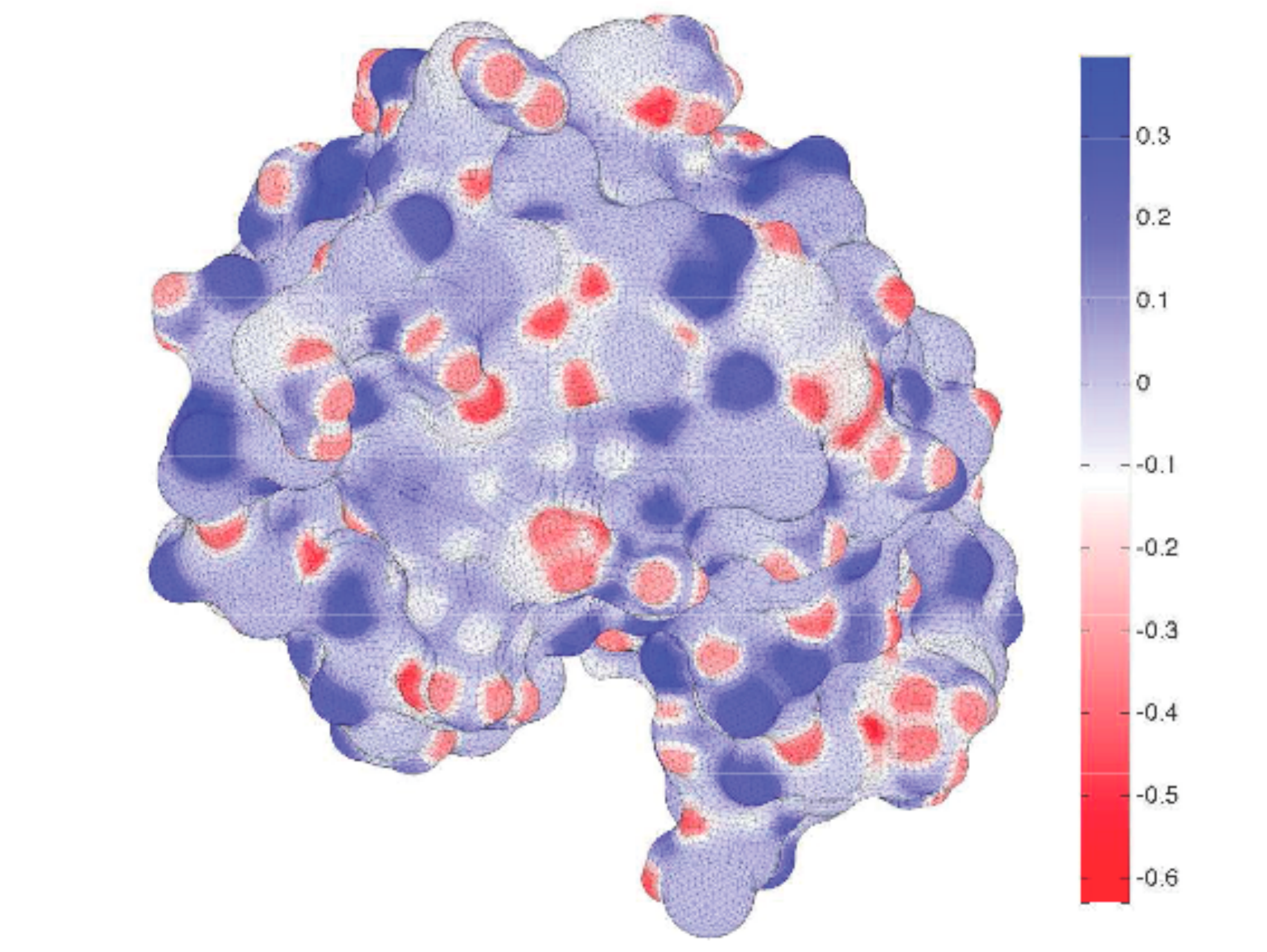}}
	\caption{The induced charge distribution for an isolated lysozyme
     molecule.  The surface has been discretized into 102,486 planar
     triangles.}
	\label{fig:lysozyme-surface}
\end{figure}

A collection of 1000 proteins is shown in
Figure~\ref{fig:lysozyme-cartoon}.  The surface charge density for an
isolated lysozyme molecule, computed using our method, is plotted in
Figure~\ref{fig:lysozyme-surface}. Of course, actually implementing a
Brownian-dynamics method requires some special adaptation of
\bem~\cite{BordnerHuber2003,LuZhangMcCammon2005}, which we have not implemented.  The
calculation did not employ periodic boundary conditions, which would
also require further adaptation of the
solver\cite{YokotaSheelObi2007}.


\begin{figure}
	\centering
	{\includegraphics[width=0.95\columnwidth]{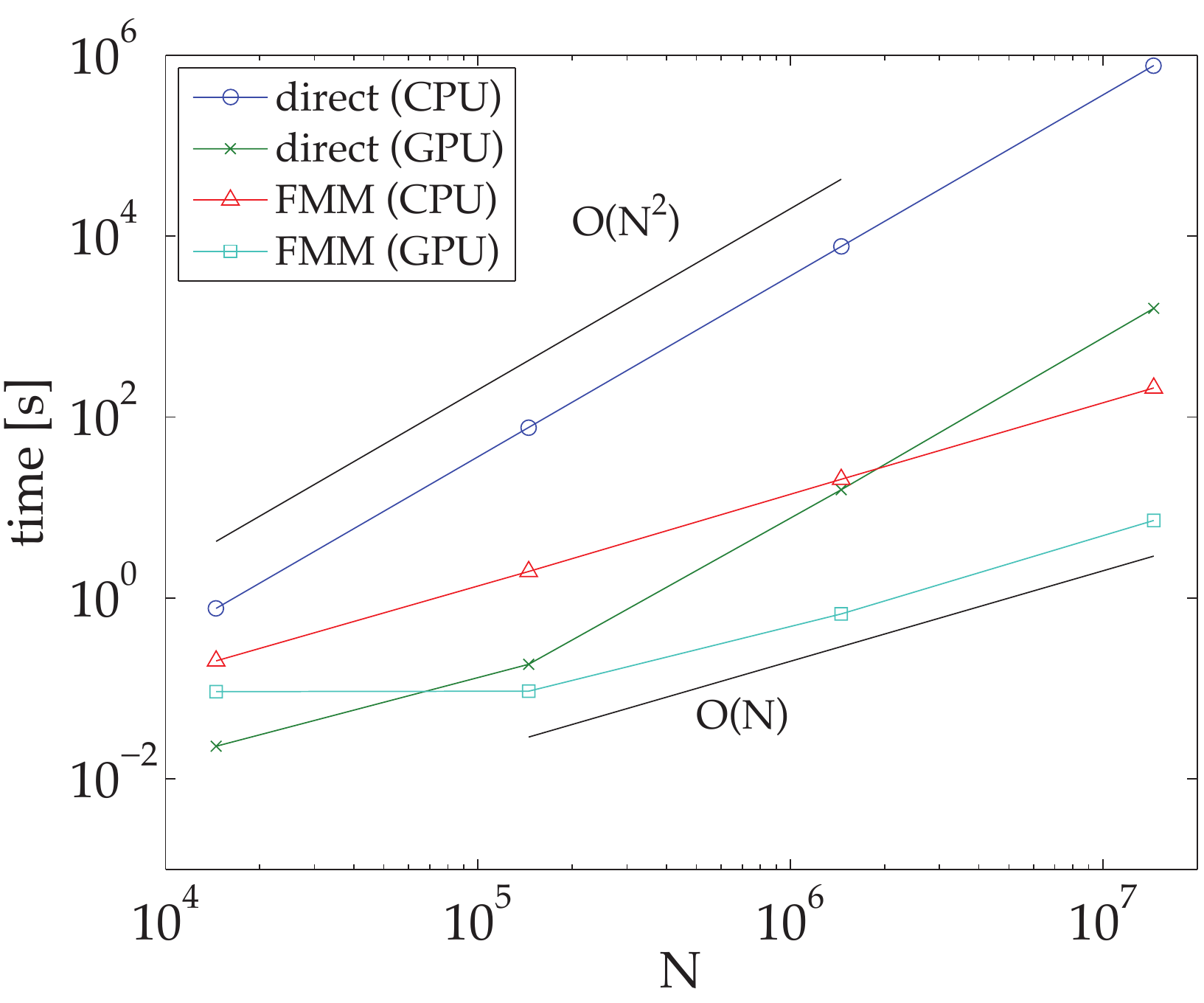}}
	\caption{Average times required to compute the dense matrix--vector
     products for the \bibee model, using a direct method or the \fmm
     on a single \gpu, as a function of the total number of boundary
     elements in lysozyme arrays; each lysozyme surface was
     discretized into 102,486 boundary elements.}
	\label{fig:1oit-timing}
\end{figure}

Figure~\ref{fig:1oit-timing} shows the computation times required for
single-\gpu\ \bibee calculations of lysozyme arrays with different
numbers of proteins.  As expected, the direct method scales
quadratically with the number of boundary elements, and the \fmm
scales linearly.  Note that \bibee \textbf{on a single \gpu\ requires less than
one second to compute electrostatics with up to one million unknowns}.
The largest simulation we have conducted consists of 10,648
molecules, where each surface was discretized into 102,486 elements.
This calculation, which models more than 20~million atoms and
possesses over \textbf{one billion unknowns}, required only about one
minute per \bem iteration on 512 nodes.

Our new ability to rigorously simulate such large systems enables the
assessment and improvement of current heuristics.  For example, one
heuristic assumes that the electrostatic potential generated by each
protein does not depend on the presence of surrounding proteins, an
assumption motivated by the high dielectric solvent, which might also
contain mobile ions.  In the boundary-integral equation formulation,
this heuristic takes the form of computing the induced surface charge
for a single isolated molecule and then employing that surface charge
for each molecule in the solution, without solving the full \bem
problem self-consistently.  Our fast solver raises the possibility of
more detailed checks of such assumptions, by enabling computationally
feasible comparison between the heuristic energy and the actual energy
computed by realizing a full \bem solution.

\section{Discussion}

We have built a boundary-element method (\bem) solver using the \bibee approximation on top of a fast-multipole method (\fmm) code with hardware acceleration using
\gpu s.  To demonstrate the solver's speed and scalability, we have
simulated several problems in electrostatics which model the
interactions between biological molecules, but it can also be used to address numerous other problems, including problems in electromagnetics and the vortex particle formulation in fluid dynamics.  The combination of optimal algorithms and modern hardware acceleration can lead to a qualitative shift in the methodology of computational protein science, and improve our ability to compute meaningfully converged quantities to be compared with experiments.

The performance demonstrated here will carry over to more sophisticated \bem applications with better accuracy, so we emphasize that the present work should be seen as a demonstration of the performance offered by the algorithms and hardware, rather than as a demonstration of the particular electrostatic energies calculated by the present implementation.
Improved methods include the use of collocation or Galerkin formulations~\cite{Bardhan2009} or curved rather than planar elements~\cite{LiangSubramaniam1997,AltmanBardhanWhiteTidor09}.  The present paper provides motivation to incorporate such methods into
our solver for fast, accurate computation of molecular
electrostatics.  As a second step beyond the implementation described here, we will extend our \fmm\ kernels in order to treat linearized Poisson--Boltzmann problems~\cite{JufferETal1991,LuETal2006}.  

In summary, substantial development is needed before the solver can be thought of as mature enough to be used for scientific studies even for simulations of fixed structures. Even with \fmm on \gpu\ it seems that the possibility of \bem-based implicit-solvent molecular dynamics remains quite distant, owing to the numerous unresolved theoretical issues.
When coupled with higher-order \bem techniques, however, the fast multipole method with \gpu\ acceleration should enable the routine calculation of quantities that were previously impractical due to computational cost.  For example, it is
now entirely feasible to conduct detailed electrostatic component
analyses of protein--protein interactions in vital biological processes.  Such studies~\cite{CarrascalGreen2010} requiring thousands or tens of thousands of simulations, have been prohibitively expensive until now.  To a large degree, the \fmm eliminates actual computation as a limiting factor; the generation of suitable surface meshes, a topic that has received much attention recently~\cite{ChengShi2009,ZhangXuBajaj2006}, appears to be one of the primary remaining challenges.  In addition, the acceleration afforded by our \fmm on \gpu\ also makes it practical
to sample many more protein conformations in accounting for
molecular flexibility~\cite{YouBashford1995,AlexovGunner1997,DysonWright2005}.  Finally, as the arrays of thousands of lysozyme molecules demonstrated,
our method could allow scientists to study complex molecular
environments with greater detail---for instance, biological solutions are quite ``crowded''~\cite{Ellis2001} and it is not yet clear how crowding impacts protein diffusion, conformation, and association~\cite{Minton00,ZimmermanMinton1993}.  The \gpu\ accelerated fast multipole method allows studying mixtures of large numbers of proteins and small molecules, and therefore to study crowding in more detail.

\bigskip

The version of the code used for this work unit is currently available in an experimental branch of the \petfmm repository.  Over time, functionality for \bem\ will be added to the main branch of \petfmm, but we make available our experimental code at the time of publication of the paper, consistent with the research group's policy of openness.  To find the repository, follow the links provided in \href{http://barbagroup.bu.edu/}{http://barbagroup.bu.edu/}.

\section{Acknowledgments}
The authors thank B. Tidor for the use of the ICE (Integrated
Continuum Electrostatics) software library and B. Roux for the use of
\textsc{charmm}. MGK supported in part by the U.S. Army Research Laboratory and the U.S. Army Research Office under contract/grant number W911NF-09-0488.  LAB
acknowledges partial support from NSF grant OCI-0946441 and from
Boston University College of Engineering.

\section*{Appendix: Structure Preparation}
All molecular structures were downloaded from the Protein Data
Bank~\cite{Berman00} (\cdk 2--inhibitor accession code 1OIT; hen egg-white
lysozyme accession code 1HEL).  The \textsc{psfgen} module of
\textsc{vmd}~\cite{HumphreyETal1996} was used to add missing atoms as well as
appropriate patches at the N- and C-termini of all peptide chains.
Atomic radii and partial charges were taken from the \textsc{parse} parameter
set~\cite{SitkoffShartHonig1994}.  All surface meshes were generated using
\textsc{msms}~\cite{MSMS}.

\bibliographystyle{plain}

\begin{thebibliography}{10}

\bibitem{AlexovGunner1997}
E.~G. Alexov and M.~R. Gunner.
\newblock Incorporating protein conformational flexibility into the calculation
  of {pH}-dependent protein properties.
\newblock {\em Biophys.\ J.}, 72(5):2075--2093, 1997.

\bibitem{AltmanETal2006}
M.~D. Altman, J.~P. Bardhan, B.~Tidor, and J.~K. White.
\newblock {FFTSVD}: A fast multiscale boundary-element method solver suitable
  for {BioMEMS} and biomolecule simulation.
\newblock {\em IEEE Trans.\ Comput.\ Aid.\ D.}, 25:274--284, 2006.

\bibitem{AltmanBardhanWhiteTidor09}
M.~D. Altman, J.~P. Bardhan, J.~K. White, and B.~Tidor.
\newblock Accurate solution of multi-region continuum electrostatic problems
  using the linearized {Poisson}--{Boltzmann} equation and curved boundary
  elements.
\newblock {\em J. Comput. Chem.}, 30:132--153, 2009.

\bibitem{AnandakrishnanETal2010}
R.~Anandakrishnan, T.~R.W. Scogland, A.~T. Fenley, J.~C. Gordon, W.-C. Feng,
  and A.~V. Onufriev.
\newblock Accelerating electrostatic surface potential calculation with
  multi-scale approximation on graphics processing units.
\newblock {\em J. Mol.\ Graph.\ Model.}, 28(8):904--910, 2010.

\bibitem{Anderson03}
M.~Anderson, J.~Beattie, G.~Breault, J.~Breed, K.~Byth, J.~Culshaw, R.~Ellston,
  S.~Green, C.~Minshull, R.~Norman, R.~Pauptit, J.~Stanway, A.~Thomas, and
  P.~Jewsbury.
\newblock Imidazo[1,2-a]pyridines: A potent and selective class of cyclin
  dependent kinase inhibitors identified through structure-based hybridization.
\newblock {\em Bioorg. Med. Chem. Lett.}, 13:3021, 2003.

\bibitem{Atkinson97}
K.~E. Atkinson.
\newblock {\em The Numerical Solution of Integral Equations of the Second
  Kind}.
\newblock Cambridge University Press, 1997.

\bibitem{BakerETal2001}
N.~A. Baker, D.~Sept, M.~J. Holst, and J.~A. Mc{C}ammon.
\newblock Electrostatics of nanoysystems: Application to microtubules and the
  ribosome.
\newblock {\em P.\ Natl.\ Acad.\ Sci. USA}, 98:10037--10041, 2001.

\bibitem{Bardhan08_BIBEE}
J.~P. Bardhan.
\newblock Interpreting the {Coulomb}-field approximation for
  {Generalized}-{Born} electrostatics using boundary-integral equation theory.
\newblock {\em J. Chem. Phys.}, 129(144105), 2008.

\bibitem{Bardhan2009}
J.~P. Bardhan.
\newblock Numerical solution of boundary-integral equations for molecular
  electrostatics.
\newblock {\em J. Chem.\ Phys.}, 130:094102, 2009.

\bibitem{BardhanETal2007}
J.~P. Bardhan, M.~D. Altman, J.~K. White, and B.~Tidor.
\newblock Numerical integration techniques for curved-panel discretizations of
  molecule--solvent interfaces.
\newblock {\em J. Chem.\ Phys.}, 127:014701, 2007.

\bibitem{Bardhan09_PRE}
J.~P. Bardhan, R.~S. Eisenberg, and D.~Gillespie.
\newblock Discretization of the induced-charge boundary integral equation.
\newblock {\em Phys. Rev. E}, 80(011906), 2009.

\bibitem{BardhanKnepleyAnitescu09}
J.~P. Bardhan, M.~G. Knepley, and M.~Anitescu.
\newblock Bounding the electrostatic free energies associated with linear
  continuum models of molecular solvation.
\newblock {\em J. Chem.\ Phys.}, 130(10):104108, 2009.

\bibitem{Berman00}
H.~M. Berman, J.~Westbrook, Z.~Feng, G.~Gilliland, T.~N. Bhat, H.~Weissig,
  I.~N. Shindyalov, and P.~E. Bourne.
\newblock The protein data bank.
\newblock {\em Nucleic Acids Res.}, 28(1):235--242, 2000.

\bibitem{BharadwajETal1995}
R.~Bharadwaj, A.~Windemuth, S.~Sridharan, B.~Honig, and A.~Nicholls.
\newblock The fast multipole boundary element method for molecular
  electrostatics: An optimal approach for large systems.
\newblock {\em J. Comput.\ Chem.}, 16(7):898--913, 1995.

\bibitem{BordnerHuber2003}
A.~J. Bordner and G.~A. Huber.
\newblock Boundary element solution of the linear {Poisson--Boltzmann} equation
  and a multipole method for the rapid calculation of forces on macromolecules
  in solution.
\newblock {\em J. Comput.\ Chem.}, 24(3):353--367, 2003.

\bibitem{Borgis03}
D.~Borgis, N.~L\'{e}vy, and M.~Marchi.
\newblock Computing the electrostatic free-energy of complex molecules: the
  variational {Coulomb} field approximation.
\newblock {\em J. Chem. Phys.}, 119(6):3516--3528, 2003.

\bibitem{BrooksETal1983}
B.~R. Brooks, R.~E. Bruccoleri, B.~D. Olafson, D.~J. States, S.~Swaminathan,
  and M.~Karplus.
\newblock {CHARMM:} {A} program for macromolecular energy, minimization, and
  dynamics calculations.
\newblock {\em J. Comput.\ Chem.}, 4:187--217, 1983.

\bibitem{BruccoleriETal1997}
R.~E. Bruccoleri, J.~Novotny, M.~E. Davis, and K.~A. Sharp.
\newblock Finite difference {Poisson}--{Boltzmann} electrostatic calculations:
  increased accuracy achieved by harmonic dielectric smoothing and charge
  antialiasing.
\newblock {\em J. Comput.\ Chem.}, 18(2):268--276, 1997.

\bibitem{Caravella99}
J.~A. Caravella, Jeffrey~D. Carbeck, D.~C. Duffy, G.~M. Whitesides, and
  B.~Tidor.
\newblock Long-range electrostatic contributions to protein-ligand binding
  estimated using protein charge ladders, affinity capillary electrophoresis,
  and continuum electrostatic theory.
\newblock {\em J. Am. Chem. Soc.}, 121:4340--4347, 1999.

\bibitem{CarrascalGreen2010}
N.~Carrascal and D.~F. Green.
\newblock Energetic decomposition with the generalized-{Born} and
  {Poisson}-boltzmann solvent models: Lessons from association of {G}-protein
  components.
\newblock {\em J. Phys.\ Chem.\ B}, 114(15):5096--5116, 2010.
\newblock PMID: 20355699.

\bibitem{Chang00}
R.~C. Chang, D.~Asthagiri, and A.~M. Lenhoff.
\newblock Measured and calculated effects of mutations in bacteriophage {T4}
  lysozyme on interactions in solution.
\newblock {\em Proteins}, 41:123--132, 2000.

\bibitem{ChengShi2009}
H.-L. Cheng and X.~Shi.
\newblock Quality mesh generation for molecular skin surfaces using restricted
  union of balls.
\newblock {\em Comp. Geom.-Theor. Appl.}, 42(3):196--206, 2009.

\bibitem{Chipman97}
D.~M. Chipman.
\newblock Charge penetration in dielectric models of solvation.
\newblock {\em J. Chem. Phys.}, 106:10194--10206, 1997.

\bibitem{Connolly1983a}
M.~L. Connolly.
\newblock {Analytical} molecular surface calculation.
\newblock {\em J. Appl. Cryst.}, 16:548--558, 1983.

\bibitem{CruzKnepleyBarba2010}
F.~A. Cruz, M.~G. Knepley, and L.~A. Barba.
\newblock {PetFMM}\,----a dynamically load-balancing parallel fast multipole
  library.
\newblock {\em Int.\ J. Num.\ Meth.\ Engineering}, 85(4):403--428, 2010.

\bibitem{DardenETal1993}
T.~Darden, D.~York, and L.~Pedersen.
\newblock Particle mesh {Ewald}: An {$N \log N$} method for {Ewald} sums in
  large systems.
\newblock {\em J. Chem.\ Phys.}, 93:10089--10092, 1993.

\bibitem{DijkstraMattheij06}
W.~Dijkstra and R.~M.~M. Mattheij.
\newblock The condition number of the {BEM}-matrix arising from {Laplace's}
  equation.
\newblock Technical report, OAI Repository of the Technische Universiteit
  Eindhoven (TU/e) [http://cache.libr.tue.nl:1972/csp/dare/DARE.Repository.cls]
  (Netherlands), 2006.

\bibitem{DysonWright2005}
H.~J. Dyson and P.~E. Wright.
\newblock Intrinsically unstructured proteins and their functions.
\newblock {\em Nat. Rev. Mol. Cell. Biol.}, 6:197--208, 2005.

\bibitem{Elcock04}
A.~H. Elcock.
\newblock Molecular simulations of diffusion and association in
  multimacromolecular systems.
\newblock {\em Meth. Enzymol.}, 383:166--198, 2004.

\bibitem{Ellis2001}
R.~J. Ellis.
\newblock Macromolecular crowding: an important but neglected aspect of the
  intracellular environment.
\newblock {\em Curr.\ Opin.\ Chem.\ Biol.}, 11(1):114--119, 2001.

\bibitem{FanETal2005}
H.~Fan, A.~E. Mark, J.~Zhu, and B.~Honig.
\newblock Comparative study of generalized {Born} models: protein dynamics.
\newblock {\em P. Natl. Aca. Sci. USA}, 102(19):6760--6764, 2005.

\bibitem{FenleyETal2008}
A.~T. Fenley, J.~C. Gordon, and A.~Onufriev.
\newblock An analytical approach to computing biomolecular electrostatic
  potential. {I}. derivation and analysis.
\newblock {\em J. Chem.\ Phys.}, 129(7):075101, 2008.

\bibitem{Gabdoulline97}
R.~R. Gabdoulline and R.~C. Wade.
\newblock Simulation of the diffusional association of barnase and barstar.
\newblock {\em Biophys. J.}, 72:1917--1929, 1997.

\bibitem{GengYuWei2007}
W.~H. Geng, S.~N. Yu, and G.~W. Wei.
\newblock Treatment of charge singularities in implicit solvent models.
\newblock {\em J. Chem.\ Phys.}, 127:114106, 2007.

\bibitem{Ghosh98}
A.~Ghosh, C.~S. Rapp, and R.~A. Friesner.
\newblock Generalized {B}orn model based on a surface integral formulation.
\newblock {\em J. Phys. Chem. B}, 102:10983--10990, 1998.

\bibitem{GilsonMcCammonMadura1995}
M.~K. Gilson, J.~A. McCammon, and J.~D. Madura.
\newblock Molecular-dynamics simulation with a continuum electrostatic
  representation of the solvent.
\newblock {\em J. Comput.\ Chem.}, 16(9):1081--1095, 1995.

\bibitem{GilsonETal1985}
M.~K. Gilson, A.~Rashin, R.~Fine, and B.~Honig.
\newblock On the calculation of electrostatic interactions in proteins.
\newblock {\em J. Mol.\ Biol.}, 184:503--516, 1985.

\bibitem{GreenTidor2005}
D.~F. Green and B.~Tidor.
\newblock Design of improved protein inhibitors of {HIV}-1 cell entry:
  Optimization of electrostatic interactions at the binding interface.
\newblock {\em Proteins}, 60:644--657, 2005.

\bibitem{GreengardRokhlin1987}
L.~Greengard and V.~Rokhlin.
\newblock A fast algorithm for particle simulations.
\newblock {\em J. Comput.\ Phys.}, 73(2):325--348, 1987.

\bibitem{HamadaNarumiYokotaYasuokaNitadoriTaiji09}
T.~Hamada, T.~Narumi, R.~Yokota, K.~Yasuoka, K.~Nitadori, and M.~Taiji.
\newblock 42 {TFlops} hierarchical {N}-body simulations on {GPUs} with
  applications in both astrophysics and turbulence.
\newblock In {\em SC '09: Proceedings of the Conference on High Performance
  Computing Networking, Storage and Analysis}, pages 1--12, New York, NY, 2009.
  ACM.

\bibitem{HardyStoneSchulten2009}
D.~J. Hardy, J.~E. Stone, and K.~Schulten.
\newblock Multilevel summation of electrostatic potentials using graphics
  processing units.
\newblock {\em Parallel Comput.}, 35:164--177, 2009.

\bibitem{Hess62}
J.~L. Hess and A.~M.~O. Smith.
\newblock Calculation of non-lifting potential flow about arbitrary
  three-dimensional bodies.
\newblock {\em J. Ship Res.}, 8(2):22--44, 1962.

\bibitem{HumphreyETal1996}
W.~Humphrey, A.~Dalke, and K.~Schulten.
\newblock {VMD}: Visual molecular dynamics.
\newblock {\em J. Mol.\ Graphics}, 14(1):33--38, 1996.

\bibitem{Jorgensen04}
W.~L. Jorgensen, J.~P. Ulmschneider, and J.~{Tirado-Rives}.
\newblock Free energies of hydration from a generalized {Born} model and an
  all-atom force field.
\newblock {\em J. Phys. Chem. B}, 108:16264--16270, 2004.

\bibitem{JufferETal1991}
A.~H. Juffer, E.~F.~F. Botta, B.~A.~M. van Keulen, A.~van~der Ploeg, and
  H.~J.~C. Berendsen.
\newblock The electric potential of a macromolecule in a solvent: A fundamental
  approach.
\newblock {\em J. Comp.\ Phys.}, 97(1):144--171, 1991.

\bibitem{KuoETal2002}
S.~S. Kuo, M.~D. Altman, J.~P. Bardhan, B.~Tidor, and J.~K. White.
\newblock Fast methods for simulation of biomolecule electrostatics.
\newblock In {\em Proceedings of the 2002 IEEE/ACM international conference on
  Computer-aided design}, ICCAD '02, pages 466--473, New York, NY, USA, 2002.
  ACM.

\bibitem{LashukETal2009}
I.~Lashuk, A.~Chandramowlishwaran, H.~Langston, T.~Nguyen, R.~Sampath,
  A.~Shringarpure, R.~Vuduc, L.~Ying, D.~Zorin, and G.~Biros.
\newblock A massively parallel adaptive fast-multipole method on heterogeneous
  architectures.
\newblock In {\em Proceedings of the Conference on High Performance Computing
  Networking, Storage and Analysis, SC '09}, pages 1--12, Portland, Oregon,
  November 2009.

\bibitem{LiangSubramaniam1997}
J.~Liang and S.~Subramaniam.
\newblock Computation of molecular electrostatics with boundary element
  methods.
\newblock {\em Biophys.\ J.}, 73(4):1830--1841, 1997.

\bibitem{LiuGrinterZou2009}
H.-Y. Liu, S.~Z. Grinter, and X.~Zou.
\newblock Multiscale generalized {Born} modeling of ligand binding energies for
  virtual database screening.
\newblock {\em J. Phys.\ Chem.\ B}, 113(35):11793--11799, 2009.
\newblock PMID: 19678651.

\bibitem{LuETal2006}
B.~Lu, X.~Cheng, J.~Huang, and J.~A. McCammon.
\newblock Order {$N$} algorithm for computation of electrostatic interactions
  in biomolecular systems.
\newblock {\em P.\ Natl.\ Acad.\ Sci. USA}, 103(51):19314--19319, 2006.

\bibitem{LuMcCammon2007}
B.~Lu and J.~A. McCammon.
\newblock Improved boundary element methods for {Poisson}--{Boltzmann}
  electrostatic potential and force calculations.
\newblock {\em J. Chem.\ Theory Comput.}, 3:1134--1142, 2007.

\bibitem{LuZhangMcCammon2005}
B.~Lu, D.~Zhang, and J.~A. McCammon.
\newblock Computation of electrostatic forces between solvated molecules
  determined by the {Poisson}--{Boltzmann} equation using a boundary element
  method.
\newblock {\em J. Chem.\ Phys.}, 122(21):214102, 2005.

\bibitem{MassovaKollman2000}
I.~Massova and P.~A. Kollman.
\newblock Combined molecular mechanical and continuum solvent approach
  {MM-PBSA/GBSA} to predict ligand binding.
\newblock {\em Perspect.\ Drug Discov.}, 18:113--135, 2000.

\bibitem{McGuffee06}
S.~R. {McGuffee} and A.~H. Elcock.
\newblock Atomistically detailed simulations of concentrated protein solutions:
  the effects of salt, {pH}, point mutations, and protein concentration in
  simulations of 1000-molecule systems.
\newblock {\em J. Am. Chem. Soc.}, 128:12098--12110, 2006.

\bibitem{Miertus1981}
S.~Miertus, E.~Scrocco, and J.~Tomasi.
\newblock Electrostatic interactions of a solute with a continuum -- a direct
  utilization of \textit{ab initio} molecular potentials for the prevision of
  solvent effects.
\newblock {\em Chem. Phys.}, 55(1):117--129, 1981.

\bibitem{Minton00}
A.~P. Minton.
\newblock Implications of macromolecular crowding for protein assembly.
\newblock {\em Curr. Opin. Struc. Biol.}, 10:34--39, 2000.

\bibitem{NaborsWhite1991}
K.~Nabors and J.~White.
\newblock {FastCap}: A multipole accelerated 3-{D} capacitance extraction
  program.
\newblock {\em IEEE Trans.\ Computer-Aided Design}, 10(11):1447--1459, 1991.

\bibitem{Neal95}
B.~L. Neal and A.~M. Lenhoff.
\newblock Excluded volume contribution to the osmotic second virial coefficient
  for proteins.
\newblock {\em AIChE J.}, 41:1010--1014, 1995.

\bibitem{Newman86}
J.~N. Newman.
\newblock Distribution of sources and normal dipoles over a quadrilateral
  panel.
\newblock {\em J. Eng. Math.}, 20(2):113--126, 1986.

\bibitem{cuda-guide}
{NVIDIA Corp.}
\newblock {CUDA} programming guide version 2.2.1, May 2009.

\bibitem{PhillipsETal2005}
J.~C. Phillips, R.~Braun, W.~Wang, J.~Gumbart, E.~Tajkhorshid, E.~Villa,
  C.~Chipot, R.~D. Skeel, L.~Kale, and K.~Schulten.
\newblock Scalable molecular dynamics with {NAMD}.
\newblock {\em J. Comput.\ Chem.}, 26:1781--1802, 2005.

\bibitem{PhillipsWhite1997}
J.~R. Phillips and J.~K. White.
\newblock A precorrected-{FFT} method for electrostatic analysis of complicated
  3-{D} structures.
\newblock {\em IEEE Trans.\ Comput.\ Aid.\ D.}, 16(10):1059--1072, 1997.

\bibitem{Qiu97}
D.~Qiu, P.~S. Shenkin, F.~P. Hollinger, and W.~C. Still.
\newblock The {GB/SA} continuum model for solvation. {A} fast analytical method
  for the calculation of approximate {Born} radii.
\newblock {\em J. Phys. Chem. A}, 101(16):3005--3014, 1997.

\bibitem{RahimianETal2010}
A.~Rahimian, I.~Lashuk, S.~Veerapaneni, A.~Chandramowlishwaran, D.~Malhotra,
  L.~Moon, R.~Sampath, A.~Shringarpure, J.~Vetter, R.~Vuduc, D.~Zorin, and
  G.~Biros.
\newblock Petascale direct numerical simulation of blood flow on 200k cores and
  heterogeneous architectures.
\newblock In {\em Proceedings of the 2010 ACM/IEEE International Conference for
  High Performance Computing, Networking, Storage and Analysis}, SC '10, pages
  1--11, Washington, DC, USA, 2010. IEEE Computer Society.

\bibitem{Richards1977a}
F.~M. Richards.
\newblock {Areas,} volumes, packing, and protein structure.
\newblock {\em Ann. Rev. Biophys. Bioeng.}, 6:151--176, 1977.

\bibitem{Rokhlin1983}
V.~Rokhlin.
\newblock Rapid solution of integral equation of classical potential theory.
\newblock {\em J. Comput.\ Phys.}, 60:187--207, 1983.

\bibitem{Romanov04}
A.~N. Romanov, S.~N. Jabin, Y.~B. Martynov, A.~V. Sulimov, F.~V. Grigoriev, and
  V.~B. Sulimov.
\newblock Surface generalized {Born} method: {A} simple, fast, and precise
  implicit solvent model beyond the {Coulomb} approximation.
\newblock {\em J. Phys. Chem. A}, 108(43):9323--9327, 2004.

\bibitem{Rush66}
S.~Rush, A.~H. Turner, and A.~H. Cherin.
\newblock Computer solution for time-invariant electric fields.
\newblock {\em J. Appl. Phys.}, 37(6):2211--2217, 1966.

\bibitem{SaadSchultz1986}
Y.~Saad and M.~Schultz.
\newblock {GMRES}: A generalized minimal residual algorithm for solving
  nonsymmetric linear systems.
\newblock {\em SIAM J. Sci.\ Stat.\ Comput.}, 7:856--869, 1986.

\bibitem{MSMS}
M.~F. Sanner.
\newblock Molecular surface computation home page.
\newblock http://mgltools.scripps.edu/packages/MSMS (last checked Feb.\ 4,
  2010), 1996.

\bibitem{SchutzWarshel2001}
C.~N. Schutz and A.~Warshel.
\newblock What are the dielectric constants of proteins and how to validate
  electrostatic models?
\newblock {\em Proteins}, 44:400--417, 2001.

\bibitem{SharpHonig1990b}
K.~A. Sharp and B.~Honig.
\newblock Calculating total electrostatic energies with the nonlinear
  {Poisson}--{Boltzmann} equation.
\newblock {\em J. Phys.\ Chem.}, 94(19):7684--7692, 1990.

\bibitem{SharpHonig1990}
K.~A. Sharp and B.~Honig.
\newblock Electrostatic interactions in macromolecules: Theory and
  applications.
\newblock {\em Ann.\ Rev.\ Biophys.\ Bio.}, 19:301--332, June 1990.

\bibitem{Shaw85}
P.~B. Shaw.
\newblock Theory of the {Poisson Green's}-function for discontinuous dielectric
  media with an application to protein biophysics.
\newblock {\em Phys. Rev. A}, 32(4):2476--2487, 1985.

\bibitem{SitkoffShartHonig1994}
D.~Sitkoff, K.~A. Sharp, and B.~Honig.
\newblock Accurate calculation of hydration free energies using macroscopic
  solvent models.
\newblock {\em J. Phys.\ Chem.}, 98(7):1978--1988, 1994.

\bibitem{SkeelETal2002}
R.~D. Skeel, I.~Tezcan, and D.~J. Hardy.
\newblock Multiple grid methods for classical molecular dynamics.
\newblock {\em J. Comput.\ Chem.}, 23(6):673--684, 2002.

\bibitem{Spector00}
S.~Spector, M.~H. Wang, S.~A. Carp, J.~Robblee, Z.~S. Hendsch, R.~Fairman,
  B.~Tidor, and D.~P. Raleigh.
\newblock {Rational} modification of protein stability by the mutation of
  charged surface residues.
\newblock {\em Biochemistry}, 39:872--879, 2000.

\bibitem{StillETal1990}
W.C. Still, A.~Tempczyk, R.~C. Hawley, and T.~F. Hendrickson.
\newblock Semianalytical treatment of solvation for molecular mechanics and
  dynamics.
\newblock {\em J. Am.\ Chem.\ Soc.}, 112(16):6127--6129, 1990.

\bibitem{StorkTavan2007}
M.~Stork and P.~Tavan.
\newblock Electrostatics of proteins in dielectric solvent continua. {I.
  Newton's} third law marries {qE} forces.
\newblock {\em J. Chem.\ Phys.}, 126(16):165105, 2007.

\bibitem{TakahashiHamada2009}
T.~Takahashi and T.~Hamada.
\newblock {GPU}-accelerated boundary element method for {Helmholtz'} equation
  in three dimensions.
\newblock {\em Int.\ J. Num.\ Meth.\ Eng.}, 80(10):1295--1321, 2009.

\bibitem{VizcarraMayo2005}
C.~L. Vizcarra and S.~L. Mayo.
\newblock Electrostatics in computational protein design.
\newblock {\em Curr.\ Opin.\ Chem.\ Biol.}, 9(6):622--626, 2005.

\bibitem{WarrenSalmon1993}
M.~S. Warren and J.~K. Salmon.
\newblock A parallel hashed oct-tree {N}-body algorithm.
\newblock In {\em Proceedings of the 1993 ACM/IEEE Conference on
  Supercomputing}, pages 12--21, New York, 1993. ACM.

\bibitem{WarshelETal2006}
A.~Warshel, P.~K. Sharma, M.~Kato, and W.~W. Parson.
\newblock Modeling electrostatic effects in proteins.
\newblock {\em Biochim.\ Biophys.\ Acta}, 1764:1647--1676, 2006.

\bibitem{WarwickerWatson1982}
J.~Warwicker and H.~C. Watson.
\newblock Calculation of the electric potential in the active site cleft due to
  alpha-helix dipoles.
\newblock {\em J. Mol.\ Biol.}, 157:671--679, 1982.

\bibitem{YingBirosZorin2004}
L.~Ying, G.~Biros, and D.~Zorin.
\newblock A kernel-independent adaptive fast multipole algorithm in two and
  three dimensions.
\newblock {\em J. Comput.\ Phys.}, 196(2):591--626, 2004.

\bibitem{YokotaSheelObi2007}
R.~Yokota, T.~K. Sheel, and S.~Obi.
\newblock Calculation of isotropic turbulence using a pure {Lagrangian} vortex
  method.
\newblock {\em J. Comput.\ Phys.}, 226(2):1589--1606, 2007.

\bibitem{Yoon90}
B.~J. Yoon and A.~M. Lenhoff.
\newblock A boundary element method for molecular electrostatics with
  electrolyte effects.
\newblock {\em J. Comput. Chem.}, 11(9):1080--1086, 1990.

\bibitem{YouBashford1995}
T.~J. You and D.~Bashford.
\newblock Conformation and hydrogen ion titration of proteins: a continuum
  electrostatic model with conformational flexibility.
\newblock {\em Biophys.\ J.}, 69(5):1721--1733, 1995.

\bibitem{ZhangXuBajaj2006}
Y.~Zhang, G.~Xu, and C.~Bajaj.
\newblock Quality meshing of implicit solvation models of biomolecular
  structures.
\newblock {\em Comput.\ Aided Geom.\ D.}, 23(6):510--530, 2006.

\bibitem{Zhou1993}
H.~X. Zhou.
\newblock Boundary-element solution of macromolecular
  electrostatics---interaction energy between 2 proteins.
\newblock {\em Biophys.\ J.}, 65:955--963, 1993.

\bibitem{ZimmermanMinton1993}
S.~B. Zimmerman and A.~P. Minton.
\newblock Macromolecular crowding: Biochemical, biophysical, and physiological
  consequences.
\newblock {\em Ann.\ Rev.\ Biophys.\ Biomol.\ Struct.}, 22(1):27--65, 1993.

\end{thebibliography}

\end{document}